\documentclass{agujournal}

\journalname{JGR-Space Physics}
\usepackage{color}

\begin{document}

\title{Non-Maxwellianity of electron distributions near Earth's magnetopause}

%% ------------------------------------------------------------------------ %%
%
%  AUTHORS AND AFFILIATIONS
%
%% ------------------------------------------------------------------------ %%

\authors{D. B. Graham \affil{1}, Yu. V. Khotyaintsev \affil{1},  M. Andr\'{e} \affil{1}, A. Vaivads \affil{2}, 
A. Chasapis \affil{3}, W. H. Matthaeus \affil{4}, A. Retino \affil{5}, F. Valentini \affil{6}, D. J. Gershman \affil{7,8}}

\affiliation{1}{Swedish Institute of Space Physics, Uppsala, Sweden.}
\affiliation{2}{Space and Plasma Physics, School of Electrical Engineering and Computer Science, KTH Royal Institute of Technology, Stockholm, Sweden.}
\affiliation{3}{Laboratory of Atmospheric and Space Physics, University of Colorado, Boulder, CO, USA.}
\affiliation{4}{Department of Physics and Astronomy, University of Delaware, Newark, DE, USA. }
\affiliation{5}{Laboratoire de Physique des Plasmas, CNRS/Ecole Polytechnique/Sorbonne Universit\'{e}/ Universit\'{e} Paris-Sud/Observatoire de Paris, Paris, France.}
\affiliation{6}{Dipartimento di Fisica, Universit\`{a} della Calabria, Arcavacata di Rende, Italy.}
\affiliation{7}{NASA Goddard Space Flight Center, Greenbelt, MD, USA.}
\affiliation{8}{Department of Astronomy, University of Maryland, College Park, MD, USA.}

\correspondingauthor{D. B. Graham}{dgraham@irfu.se}

\begin{keypoints}
\item 1: Electron non-Maxwellianity is computed for 6 months of data ($\sim$85 million electron distributions). 
%\item 2: Electron non-Maxwellianity increases as density decreases due to hot and cold electron populations 
%in the magnetosphere. 
\item 2: Electron non-Maxwellianity is typically large in the magnetosphere due to hot and cold electron populations. 
\item 3: Enhanced non-Maxwellianity is found in reconnection regions, the bowshock, and magnetosheath turbulence. 
\end{keypoints}

\begin{abstract}
Plasmas in Earth's outer magnetosphere, magnetosheath, and solar wind are essentially collisionless. This means 
particle distributions are not typically in thermodynamic equilibrium and deviate significantly from Maxwellian distributions. 
The deviations of these distributions can be further enhanced by plasma processes, such as shocks, turbulence, 
and magnetic reconnection. Such distributions can be unstable to a wide variety of kinetic plasma instabilities, which in turn modify the electron distributions. In this paper the deviations of the observed electron distributions from a bi-Maxwellian distribution function is calculated and quantified using data from the Magnetospheric Multiscale (MMS) spacecraft. 
A statistical study from tens of millions of electron 
distributions shows that the primary source of the observed non-Maxwellianity are electron distributions consisting 
of distinct hot and cold components in Earth's low-density magnetosphere. This results in large non-Maxwellianities 
in at low densities. 
However, after performing a stastical study we find regions where large non-Maxwellianities are 
observed for a given density. Highly non-Maxwellian distributions are routinely found are Earth's bowshock, in Earth's outer magnetosphere, and in the electron diffusion regions of magnetic reconnection. Enhanced non-Maxwellianities are observed in the turbulent magnetosheath, but are intermittent and are not correlated with local processes. The causes of enhanced non-Maxwellianities are investigated. 
\end{abstract}

\section{Introduction} 
Many space and astrophysical plasmas are essentially collisionless so Coulomb collisions are unlikely to be efficient in keeping particle distributions close to thermal equalibrium, i.e., a Maxwellian distribution. As a result 
non-Maxwellian distributions can readily develop and are indeed frequently observed in space plasmas. 
In collisionless plasmas non-Maxwellian distributions can remain kinetically stable, which need not relax to Maxwellian distributions. However, non-Maxwellian distributions can be important source of instabilities and
can potentially generate a variety of electrostatic and electromagnetic waves. 
Plasma processes such as shocks, magnetic reconnection, and turbulence can further increase the 
deformations in the particle distributions from a Maxwellian distribution. Quantifying the deviation in particle distributions is crucial to understanding the effects of both large-scale processes, such as shocks and magnetic reconnection, 
and kinetic-scale processes, such as wave-particle interactions. 

At present various papers have considered the non-Maxwellianity of particle distributions in both simulations and 
observations \cite[]{greco2012,valentini2016,chasapis2018,perri2020,liang2020}. These studies have focused 
on plasma turbulence in Earth's magnetosheath and magnetic reconnection. 
The simulation results showed that ion non-Maxwellianity 
was spatially non-uniform and was associated with strong currents and temperature anisotropy 
\cite[]{greco2012,valentini2016}. Similarly, electron non-Maxwellianity was found to increase the electron diffusion region (EDR) and separatrices of magnetic reconnection \cite[]{liang2020}.
In kinetic simulations the background distributions, such as in modeling of 
magnetosheath turbulence and magnetic reconnection, are assumed to be Maxwellian. Such 
distributions are not necessarily valid in Earth's magnetosheath and magnetopause, where the background distributions can differ significantly from a Maxwellian distribution while remaining kinetically stable. 
Recent observations from the Magnetospheric Multiscale (MMS) spacecraft suggest that ion non-Maxwellianity is weakly correlated with the local current sheets in the turbulent magnetosheath \cite[]{perri2020}. In contrast, 
\cite{chasapis2018} found that electron non-Maxwellianity tended to increase in regions of strong currents. 
Estimates of the non-Maxwellianity of particle distributions based on recent MMS observations have only focused on very short time intervals. Thus, it is unclear if such deviations from Maxwellianity are statistically significant when compared with a large volume of data. Therefore, a statistical study of the non-Maxwellianity is required. 

The purpose of this paper is to investigate and quantify the non-Maxwellianity of electron distribution functions in the near 
Earth plasma environment, specifically near Earth's magnetopause, in the magnetosheath, and near the bowshock. In this paper we propose a measure of the deviation of the observed electron distribution function from a bi-Maxwellian distribution function, where temperature anisotropy is included. We show that statistically the non-Maxwellianity of electron distributions increases as plasma density decreases. Large deviations of observed electron distributions from a bi-Maxwellian distribution function are observed in the outer magnetosphere, in magnetic reconnection electron diffusion regions, at the bowshock, and 
intermittently in magnetosheath turbulence. The outline of this paper is as follows:
Section \ref{MMSdata} the data used is stated, section \ref{theorymethod} states the theory and methods used to calculate electron non-Maxwellianity, section \ref{statisticalresults} presents the statistical results, 
section \ref{casestudies} presents case studies of the reconnection ion and electron diffusion regions, the bowshock, and magnetosheath turbulence, and section \ref{conclusions} states the conclusions. 

\section{Data} \label{MMSdata}
In this paper we use high-resolution burst mode data from the four MMS spacecraft \cite[]{burch1}. 
We use particle distributions and moments from Fast Plasma Investigation (FPI) 
\cite[]{pollock1}. Three-dimensional electron distributions and moments are sampled every $30$~ms. The 
electron distributions are sampled over 32 energy channels ranging from $10$~eV to $30$~keV, which covers 
the thermal electron energy range in the outer magnetosphere and magnetosheath. 
%The azimuthal and polar angular spacing is $11.25^{\circ}$. 
Ion distributions and moments are sampled every $150$~ms. 
We use electric field ${\bf E}$ and magnetic field ${\bf B}$ data from the Electric field Double Probes (EDP) 
\cite[]{lindqvist1,ergun3} and 
Fluxgate Magnetometer (FGM) \cite[]{russell2}, respectively. The spacecraft potential $V_{sc}$ is computed from the Spin-plane Double Probes (SDP). 

%In this paper we use data from MMS's first magnetopause science phase, from September 2015 to March 2016. 
%Throughout this phase the Active Spacecraft Potential Control (ASPOC) was both on and off. When 
%ASPOC is on the spacecraft potential is reduced to $\sim 4$~V \cite[]{}. In this case photoelectrons from the spacecraft 
%are unlikely to be observed by FPI. However, internal photoelectrons in the FPI-DES detectors are observed at low energies \cite[]{gershman2017}.

\section{Theory and methods} \label{theorymethod}
In this section we define the non-Maxwellianity parameter $\epsilon$. 
The non-Maxwellianity $\epsilon$ is defined as 
the velocity space integrated absolute difference between the observed distribution function and a model bi-Maxwellian distribution function given by the particle moments: 
\begin{equation}
\epsilon = \frac{1}{2 n_e} \int_{v,\theta,\phi} |f_e(v,\theta,\phi) - f_{\mathrm{model}}(v,\theta,\phi)| v^2 \sin{\theta}  dv d\theta d\phi, 
\label{epsilon}
\end{equation}
where $v$ is the electron speed, $\theta$ is the polar angle, $\phi$ is the azimuthal angle, $n_e$ is the electron number density, $f_e(v,\theta,\phi)$ is the observed velocity-space density, and $f_{\mathrm{model}}(v,\theta,\phi)$ is the velocity 
space density of the model distribution. 
The factor $1/(2 n_e)$ normalizes $\epsilon$ to a dimensionless quantity with domain $\epsilon \in [0, 1]$, where $\epsilon = 0$ corresponds to no deviation of $f_e(v,\theta,\phi)$ from $f_{\mathrm{model}}(v,\theta,\phi)$ and 
$\epsilon = 1$ corresponds to a maximal deviation, such that there is no overlap $f_e(v,\theta,\phi)$ and $f_{\mathrm{model}}(v,\theta,\phi)$ in velocity space.
For $f_{\mathrm{model}}$ we use a drifting bi-Maxwellian distribution, given by: 
\begin{equation}
f_{\mathrm{model}}({\bf v}) = \frac{n_e}{\pi^{3/2} v_{e,\parallel}^3} \frac{T_{e,\parallel}}{T_{e,\perp}} \exp{\left( - \frac{(v_{\parallel} - V_{\parallel})^2}{v_{e,\parallel}^2}  - \frac{(v_{\perp,1} - V_{\perp})^2 + v_{\perp,2}^2}{ v_{e,\parallel}^2 (T_{e,\perp}/T_{e,\parallel}) } \right)},
\label{bimax}
\end{equation}
where $T_{e,\parallel}$ and $T_{e,\perp}$ are the parallel and perpendicular electron temperatures, $v_{e,\parallel} = \sqrt{2 k_B T_{e,\parallel}/m_e}$ is the parallel electron thermal speed, $V_{\parallel}$ is the bulk velocity parallel to ${\bf B}$, 
$V_{\perp}$ is the magnitude of the bulk velocity perpendicular to ${\bf B}$, 
and $k_B$ is the Boltzmann constant. 
This calculation of $\epsilon$ corresponds to the lowest order 
moment, i.e., number density, so finite $\epsilon$ will result from deviations from a bi-Maxwellian at thermal 
electron energies. 
The velocity coordinates $(v_{\parallel},v_{\perp,1},v_{\perp,2})$ used in equation (\ref{bimax}) are defined such that $v_{\parallel}$ is the 
speed along the magnetic field direction, $v_{\perp,1}$ is the speed along the perpendicular bulk velocity direction, 
and $v_{\perp,2}$ is orthogonal to $v_{\parallel}$ and $v_{\perp,1}$. The parameters $n_e$, ${\bf V}_e$, and ${\bf T}_e$, used to calculate $f_{\mathrm{model}}$ are the FPI-DES electron moments \cite[]{pollock1}. 
To compute $\epsilon$ we 
transform $f_{\mathrm{model}}$ into the same coordinate system and discretize to the same energy and angle channels as the observed $f_e$ for direct comparison. 
We note that this definition of $\epsilon$ differs from the definition used in previous studies \cite[e.g.,][]{greco2012}. The definition of \cite{greco2012} is closely related to the enstrophy of the particle distribution \cite[]{servidio2017}. 
We have used the definition in equation (\ref{epsilon}) because: 
(1) The definition used in \cite{greco2012} is not dimensionless, which is problematic when there are large changes in density, such as across the magnetopause. 
(2) We have used a drifting bi-Maxwellian 
distribution as $f_{\mathrm{model}}$, rather than an isotropic Maxwellian distribution, 
so that increases in $\epsilon$ do not simply correspond to large bulk electron flows in the spacecraft frame, 
or large temperature anisotropies, which are straightforward to obtain from the particle moments. 
In equation (\ref{bimax}) we have assumed a single perpendicular electron temperature 
$T_{e,\perp}$ meaning $f_{\mathrm{model}}$ is gyrotropic, so agyrotropic features of the observed electron distribution will not be captured by the model distribution.
Thus, agyrotropic distributions, such as those found in the electron diffusion regions of magnetic reconnection, should result in an increased $\epsilon$. 

At low electron energies, there 
are several effects that can artificially increase $\epsilon$. These include: 

(1) Spacecraft photoelectrons are detected when the Active Spacecraft Potential Control (ASPOC) \cite[]{Torkar2016} is off and the spacecraft potential is larger than $\approx 10$~eV. The energy channels affected by spacecraft photoelectrons are removed and the remaining energy channels are corrected for when calculating $\epsilon$, so the effect of spacecraft photoelectrons should be small. 

(2) Photoelectrons generated inside the electron detectors produce enhancements in the phase space density \cite[]{gershman2017}. This affects the Sunward pointing detectors, and can occur at energies 
exceeding $e V_{sc}$. 

(3) Secondary photoelectrons can occur within the detector, resulting in artifically large phase-space densities at low 
energies. These electrons can occur at energies exceeding $e V_{sc}$. 

(4) Low-energy electrons can be focused along the spin-plane and axial booms, which are positively charged \cite[e.g.,][]{toledoredondo2019}. In addition, when ASPOC is on the ion plumes are emitted from the 
spacraft, modifying the motion of low-energy electrons \cite[]{barrie2019}. 
This can distort the measured electron distribution at low energies. 

While there is a model for internal photoelectrons \cite[]{gershman2017}, which can approximately remove these effects, the other 
effects are not straightforward to remove. Therefore, we simply perform the calculation of $\epsilon$ for electron 
energies $E > 28$~eV. This corresponds to neglecting the lower 4 energy channels in the FPI-DES distribution functions for phase 1a of MMS operations when evaluating equation (\ref{epsilon}). In addition, energy channels with eV/e$< V_{sc}$ 
are removed, and the energy channels are corrected by $- e V_{sc}$ when calculating $f_{\mathrm{model}}$ 
and $\epsilon$. The spacecraft potential $V_{sc}$ is computed 
from the average probe-to-spacecraft potentials of the four 
spin-plane probes. This average probe-to-spacecraft potential was compared to the cutoff energies of photoelectrons seen in FPI-DES data to calibrate $V_{sc}$ \cite[]{graham2018b}.

\begin{figure*}[htbp!]
\begin{center}
\includegraphics[width=160mm, height=140mm]{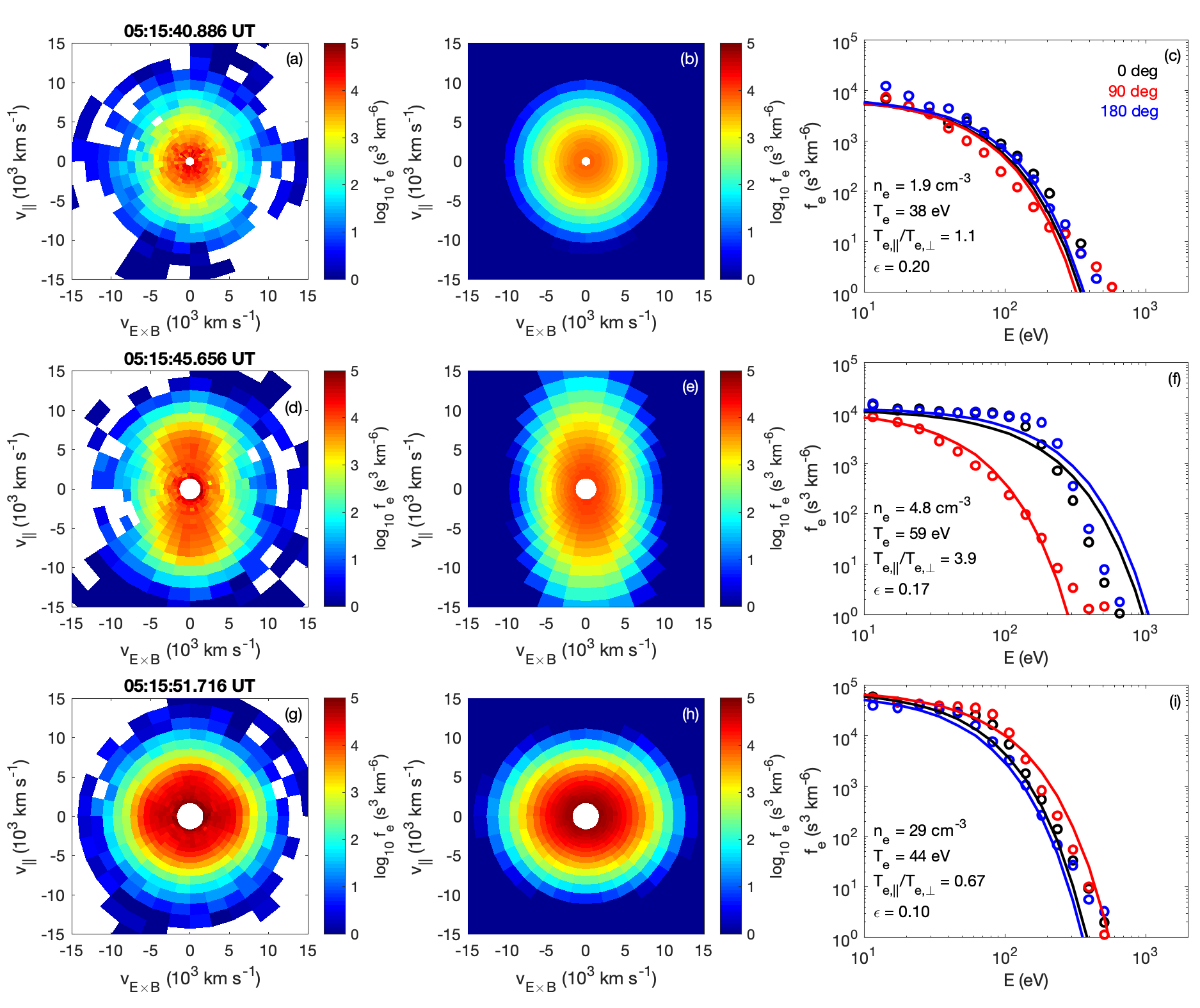}
\caption{Three examples of electron distributions and the predicted bi-Maxwellian distribution based on the electron moments from MMS1 on 30 October 2015. Panels (a), (d), and (g) show a two-dimensional slice of the observed three-dimensional electron distribution in the ${\bf B}$ and ${\bf E} \times {\bf B}$ plane. Panels (b), (e), and (h) show the modeled bi-Maxwellian distributions in the same plane. Panels (c), (f), and (i) show the phase-space densities at pitch angles $0^{\circ}$ (black), $90^{\circ}$ (red), and $180^{\circ}$ (blue) for the observed 
distributions (circles) and modeled bi-Maxwellian (solid lines). The distribution in panels (a)--(c) is in the magnetosphere, (d)--(f) is near the ion diffusion region where $T_{e,\parallel}/T_{e,\perp}$ peaks, (g)--(i) is in the 
magnetosheath where $T_{e,\parallel}/T_{e,\perp}$ is minimal. The electron distribution properties and $\epsilon$ 
of the three distributions are given in panels (c), (f), and (i).}
\label{exampledist}
\end{center}
\end{figure*}

As examples of the calculation of $\epsilon$, Figure \ref{exampledist} shows three observed electron distributions, the associated $f_{\mathrm{model}}$, and values of $\epsilon$. The distributions are from a reconnection event observed at the 
magnetopause on 30 October 2015 \cite[]{graham4}. The first distribution is in the magnetosphere close to the magnetopause 
(Figures \ref{exampledist}a--\ref{exampledist}c), the second distribution is close to the reconnection ion 
diffusion region where $T_{e,\parallel}/T_{e,\perp}$ peaks (Figures \ref{exampledist}d--\ref{exampledist}f), 
and the third distribution is in the magnetosheath where $T_{e,\parallel}/T_{e,\perp} < 1$ 
(Figures \ref{exampledist}g--\ref{exampledist}i). 

Figure \ref{exampledist}a shows an approximately isotropic electron distribution with $T_{e,\parallel}/T_{e,\perp} = 1.1$ and $\epsilon = 0.20$. 
The modeled distribution (Figure \ref{exampledist}b) is an approximately isotropic distribution. One of the 
differences between the observed and modeled distribution is the fluctuations in $f_e$ due to the counting 
statistics of the particle instrument at relatively low $n_e \lesssim 1 \, \mathrm{cm}^{-3}$, which results in an increase in $\epsilon$. This can be seen by comparing Figures \ref{exampledist}a and \ref{exampledist}b; the observed $f_e$ shows fluctuations as 
functions of speed and angle, while the modeled $f_e$ smoothly varies. Figure 
\ref{exampledist}c shows that quantitively there is some deviation of the observed distribution from $f_{\mathrm{model}}$. In particular, at pitch angle $\theta = 90^{\circ}$, the shape of $f_e$ differs from 
$f_{\mathrm{model}}$ in the thermal energy range.

For the second distribution both $f_e$ and $f_{\mathrm{model}}$ are qualitatively similar. However, Figure \ref{exampledist}f shows that there is significant deviation in $f_e$ from $f_{\mathrm{model}}$. For 
$\theta = 90^{\circ}$, $f_e$ is close to Maxwellian at thermal energies, 
while for $\theta = 0^{\circ}$ and $180^{\circ}$ a flat-top 
distribution is observed, which deviates from the shape of a Maxwellian. Thus, the flat-top distribution corresponds 
to an enhanced non-Maxwellianity of $\epsilon = 0.17$. 

The final distribution is in the high-density magnetosheath, so there is little noise in $f_e$ because the particle counts are high. Here, $f_e$ and $f_{\mathrm{model}}$ are very similar, although there are some deviations in $f_e$ from $f_{\mathrm{model}}$, as seen in Figure \ref{exampledist}i. For this distribution $\epsilon = 0.1$, 
corresponding to relatively small deviations from a bi-Maxwellian distribution. 

\section{Statistical results} \label{statisticalresults}
\begin{figure*}[htbp!]
\begin{center}
\includegraphics[width=160mm, height=45mm]{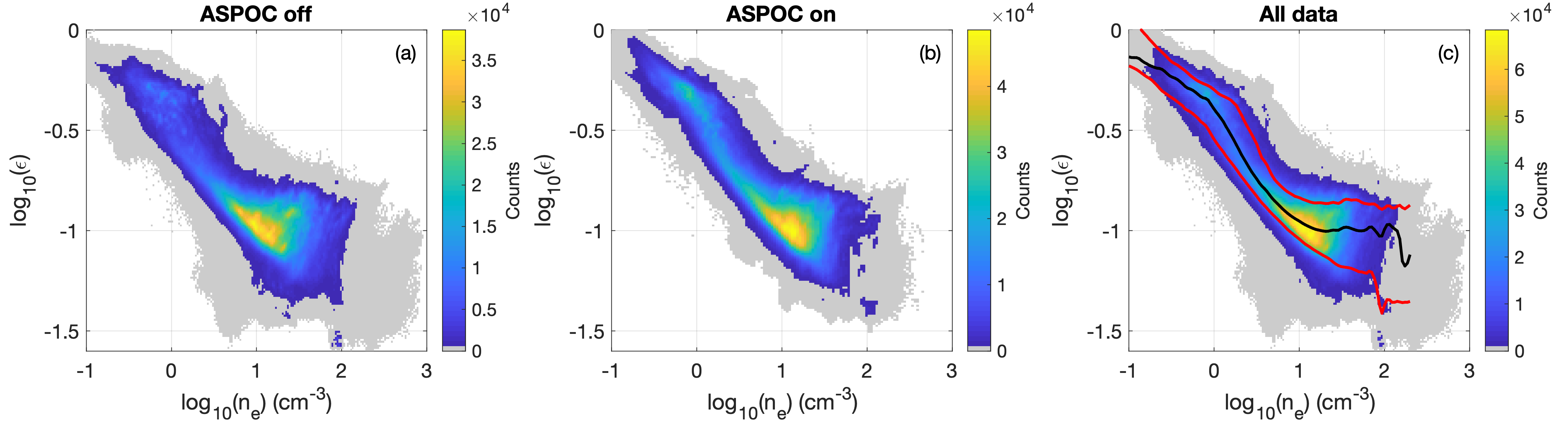}
\caption{Two-dimensional histograms of $\log_{10}(n_e)$ versus $\log_{10}(\epsilon)$. (a) Histogram for data when ASPOC is off, (b) histogram for data when ASPOC is on, and (c) histogram for all data. The color shading indicates the counts. The black line indicates the median (50th percentile) of $\epsilon$ as a function $n_e$ and the lower and upper red curves indicate the 10th and 90th percentiles as a function of $n_e$, respectively. }
\label{statsne}
\end{center}
\end{figure*}

In this section we investigate the statistical properties of $\epsilon$. We calculate $\epsilon$ for all burst mode electron data from September 2015 to March 2016 (the first magnetopause phase of the MMS mission), 
corresponding to $\sim 85$ million distributions from the four spacecraft. 
We first consider the dependence of $\epsilon$ on $n_e$. Low-densities correspond to the outer 
magnetosphere, while high densities typically correspond to the magnetosheath. 
In Figure \ref{statsne} we plot two-dimensional histograms of $\log_{10}(\epsilon)$ versus 
$\log_{10}(n_e)$ for data when ASPOC is off, ASPOC is on, and all data in panels (a)--(c), respectively. 
In all three panels we see that for $n_e \lesssim 10$~cm$^{-3}$ there is statistically an increase in $\epsilon$ as 
$n_e$ decreases, which approximately scales as $\epsilon \propto 1/\sqrt{n_e}$. For $n_e \gtrsim 10$~cm$^{-3}$, typically corresponding to the magnetosheath, we find that $\epsilon$ does not depend strongly on $n_e$. 
We find that $\epsilon$ is not strongly affected by whether or not ASPOC is on, 
except at low $n_e$, where $\epsilon$ is slightly larger when ASPOC is on. 
This might be due to the internal photoelectron emission in the FPI detectors, which is more significant at low $n_e$ and the spacecraft potential is low, or due to distortions in the observed electron distribution by the plume of ions around the spacecraft when ASPOC is on \cite[]{barrie2019}.
We do not 
find any statistical differences between the four spacecraft (not shown).  

There are two main reasons for the increase in $\epsilon$ at low $n_e$ in the magnetosphere: 

(1) In the magnetosphere distinct cold and hot electron populations are frequently present at the same time \cite[]{walsh2020}. 
In these cases 
the total effective electron temperature is 
\begin{equation}
T_e \approx \frac{n_c T_c + n_h T_h}{n_c + n_h},
\label{Teeq}
\end{equation}
where the subscripts $c$ and $h$ refer to the cold and hot electron components. When $n_c$ and $n_h$ are 
comparable $T_e$ differs significantly from $T_c$ and $T_h$, and large $\epsilon$ develop. 
In the magnetosheath the electron distributions are characterized by a single temperature, although the shape 
can differ from bi-Maxwellian distribution. This results in a smaller $\epsilon$ in the magnetosheath compared to the magnetosphere. 

(2) As $n_e$ decreases the particle counts measured in each energy and angle bin of FPI-DES decreases. 
This results in increased statistical uncertainty, corresponding to a more grainy looking distribution at lower 
density, which differs from the smooth $f_{\mathrm{model}}$ distribution, resulting in an increase in $\epsilon$. 

\begin{figure*}[htbp!]
\begin{center}
\includegraphics[width=160mm, height=90mm]{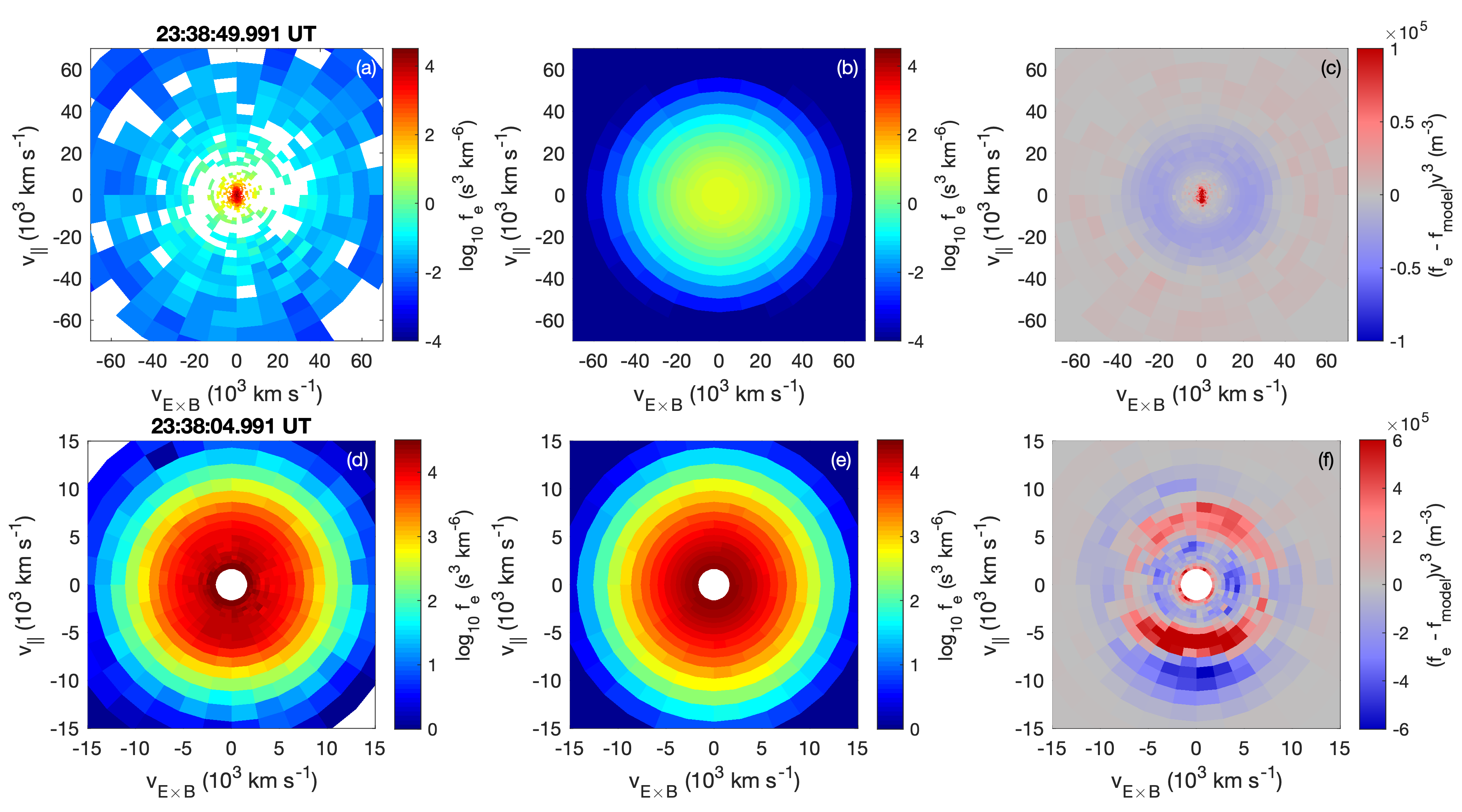}
\caption{Two examples of electron distributions and the modeled bi-Maxwellian distribution based on the electron moments from MMS1 on 06 December 2015 (cf., \cite{khotyaintsev4}). Panels (a) and (d) show a two-dimensional slice of the observed three-dimensional electron distribution in the ${\bf B}$ and ${\bf E} \times {\bf B}$ plane. 
Panels (b) and (e) show the modeled bi-Maxwellian distributions in the same plane. Panels (c) and (f) show 
$(f_e - f_{\mathrm{model}}) v^3$ in the same plane, which indicates the regions of velocity space that contribute most 
to $\epsilon$. The distribution in panels (a)--(c) is in the magnetosphere close to the magnetopause, and (d)--(f) is in the magnetosheath close to the magnetopause.}
\label{distneex}
\end{center}
\end{figure*}

The first effect iss physical and due to differences in typical magnetospheric and magnetosheath distributions, while the second effect is an instrumental effect. 
Both effects result in the statistical increase in $\epsilon$ as $n_e$ decreases, seen in Figure \ref{statsne}.  
To illustrate these two effects, we plot two electron distributions in Figure \ref{distneex}. 
The first distribution (top row) is in the magnetosphere close to the magnetopause. For this distribution 
we calculate $\epsilon = 0.69$. The second distribution 
(bottom row) is in the magnetosheath close to the magnetopause. For this distribution we calculate $\epsilon = 0.085$. Both distributions are observed by MMS1 on 06 December 2015 when the spacecraft crossed the magnetopause \cite[]{khotyaintsev4}. 
Figures \ref{distneex}a and \ref{distneex}b show 2D slices of the observed $f_e$ in the 
${\bf B}$ and ${\bf E} \times {\bf B}$ plane and $f_{\mathrm{model}}$ calculated from the particle moments. 
The distribution in Figure \ref{distneex}a is characterized by a very cold component, with $n_c = 0.3$~cm$^{-3}$ and $T_c = 27$~eV, and a hot component, with $n_h = 0.06$~cm$^{-3}$ and $T_h = 5$~keV. 
The total distribution has $n_e = 0.36$~cm$^{-3}$ and $T_e = 1$~keV. Thus, $f_{\mathrm{model}}$ shown in Figure \ref{distneex}b differs from both the cold and hot components of $f_e$, resulting in a large $\epsilon$. In Figure \ref{distneex}c we plot $(f_e - f_{\mathrm{model}}) v^3$, which 
indicates the regions of velocity space that contribute most significantly to $\epsilon$. The 
largest contribution is from low energies, where $f_e \gg f_{\mathrm{model}}$ and most of the particles are located. At intermediate energies $\sim 1$~keV, $f_e \ll f_{\mathrm{model}}$ because $f_{\mathrm{model}}$ has $T_e = 1$~keV, whereas $f_e$ is negligible due to the temperatures of the two components. 
For $E \sim 5$~keV, $f_e \gg f_{\mathrm{model}}$ due to $T_h > T_e$. As a result 
$|f_e - f_{\mathrm{model}}|$ is large over almost all velocity space making $\epsilon$ large. 

At low densities the counts per energy and angle bin are small and in many cases no counts are measured, 
as indicated by the white regions in Figure \ref{distneex}a. This results in a more grainy looking 
$f_e$, in contrast to the smooth $f_{\mathrm{model}}$ (Figure \ref{distneex}b). This results in $\epsilon$ 
tending to increase as $n_e$ decreases. At high densities the counts are very high in the thermal energy 
range (for example, the distribution in Figure \ref{distneex}d), so the effects of the finite counting statistics 
are small. In Appendix \ref{app1} we show that the most significant contribution to $\epsilon$ in the magnetosphere is the distinct cold and hot electron populations. The effect of the counting statistics on 
$\epsilon$ is smaller when hot and cold electrons are present. 

For the distribution in Figure \ref{distneex}d only a single electron population is observed, characterized 
by $n_e = 21$~cm$^{-3}$ and $T_e = 74$~eV. As a result $f_{\mathrm{model}}$ (Figure \ref{distneex}e) is very similar to $f_e$, and
$\epsilon = 0.085$. Despite $f_e$ and $f_{\mathrm{model}}$ looking similar, non-Maxwellian features are observed in the thermal energy range, as shown in Figure \ref{distneex}f. 

%This is due to the lower counting statistics as $n_e$ decreases. This effectively introduces noise into the observed electron distributions resulting in the appearance of non-Maxwellianity.

In Figure \ref{statsne}c we overplot the median $\epsilon$ ($\epsilon_{50}$) in black, and the 10th and 90th 
percentiles $\epsilon_{10}$ and $\epsilon_{90}$ (lower and upper red curves, respectively) as functions of $n_e$. We find that $\epsilon_{50} \approx 0.1$ is approximately constant for $10 \, \mathrm{cm}^{-3} \lesssim n_e 
\lesssim 100 \, \mathrm{cm}^{-3}$. Because of the strong dependence of $\epsilon$ on $n_e$, we need to consider 
the statistical median and percentiles when considering specific events to determine if the electron distributions 
are unusually non-Maxwellian compared with the median values of $\epsilon$. 
We will use these percentiles as a function of density to determine whether electron distributions in specific 
regions significantly deviate from a bi-Maxwellian distributions or not. 

\begin{figure*}[htbp!]
\begin{center}
\includegraphics[width=120mm, height=100mm]{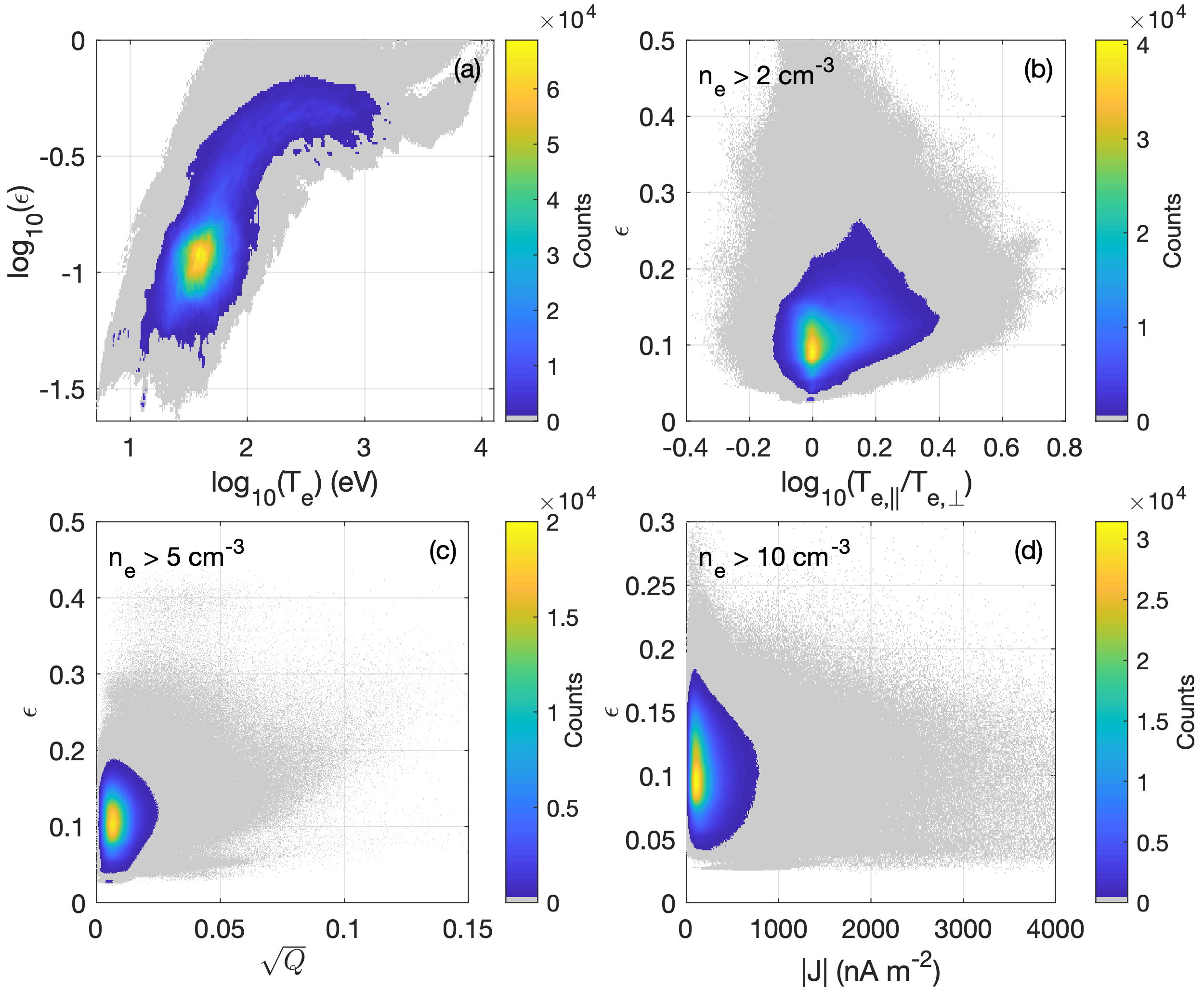}
\caption{Two-dimensional histograms of $\epsilon$ versus different plasma conditions. (a) $\log_{10}\epsilon$ versus $\log_{10}T_e$. (b) $\epsilon$  versus $\log_{10}T_{e,\parallel}/T_{e,\perp}$ for $n_e > 2$~cm$^{-3}$. 
(c) $\epsilon$ versus agyrotropy measure $\sqrt{Q}$ for $n_e > 5$~cm$^{-3}$. 
(d) $\epsilon$ versus $|{\bf J}|$ for $n_e > 2$~cm$^{-3}$.}
\label{statsprops}
\end{center}
\end{figure*}

In Figure \ref{statsprops}a we plot the histogram of $\log_{10}\epsilon$ versus $\log_{10}T_e$. We find that 
statistically $\epsilon$ increases as $T_e$ increases. This is primarily due to the statistical increase in $T_e$ 
as $n_e$ decreases. Low-density higher-temperature regions correspond to the outer magnetosphere, while 
high-density lower-temperature regions typically correspond to the magnetosheath. 
In Figure \ref{statsprops}b we plot $\epsilon$ versus $\log_{10}T_{e,\parallel}/T_{e,\perp}$. 
Overall, we find that the statistical dependence of $\epsilon$ on $T_{e,\parallel}/T_{e,\perp}$ is relatively weak. 
However, the smallest $\epsilon$ are found for $T_{e,\parallel}/T_{e,\perp} \sim 1$. In this study there are more distributions with $T_{e,\parallel}/T_{e,\perp} > 1$ than $T_{e,\parallel}/T_{e,\perp} < 1$, with a median 
and mean $T_{e,\parallel}/T_{e,\perp}$ of 1.1 and 1.2, respectively. 

We now compare $\epsilon$ with the agyrotropy of the electron distribution. We use the agyrotropy measure 
$\sqrt{Q}$ as defined in \cite{swisdak2}. This measure is based on the off-diagonal components of the electron 
pressure tensor ${\bf P}_e$ and is given by
\begin{equation}
\sqrt{Q} = \left( \frac{P_{12}^2 + P_{13}^2 + P_{23}^2}{P_{\perp}^2 + 2 P_{\perp} P_{\parallel}} \right)^{1/2}, 
\label{sqrtQ}
\end{equation}
where we have rotated the measured ${\bf P}_e$ into the field-aligned coordinates of the form: 
\begin{equation}
{\bf P}_e = 
\begin{pmatrix}
P_{\parallel} & P_{12} & P_{13} \\
P_{12} & P_{\perp} & P_{23} \\
P_{13} & P_{23} & P_{\perp}
\end{pmatrix}.
\end{equation}
The agyrotropic measure $\sqrt{Q}$ has values between $0$ and $1$, with $0$ corresponding to a gyrotropic distribution and $1$ to maximum agyrotropy. 
In Figure \ref{statsprops}c we plot the histogram of $\epsilon$ versus $\sqrt{Q}$ for $n_e > 5$~cm$^{-3}$ 
(at lower $n_e$ the agyrotropy measures tend to be unreliable). We find that the vast majority of distributions 
are approximately gyrotropic, with median and mean values of $\sqrt{Q}$ of $0.008$ and $0.009$, respectively.  
The largest values of $\sqrt{Q}$ observed are $\sim 0.1$, which correspond to values observed in the EDRs of magnetopause reconnection 
\cite[]{graham11,norgren4,webster1}. For these large values of $\sqrt{Q}$ there is a tendency of $\epsilon$ to increase with $\sqrt{Q}$. However, only a tiny fraction of the distributions have large $\sqrt{Q}$, and for most distributions there is little dependence of $\epsilon$ on $\sqrt{Q}$. This is not surprising because: (1) gyrotropic distributions can deviate significantly from a bi-Maxwellian distribution function, and (2) $\sqrt{Q}$ is based on the pressure tensor, so large large off-diagonal pressure terms may not correspond to large changes in phase-space density. 

In Figure \ref{statsprops}d we plot the histogram of $\epsilon$ versus the magnitude of the current density 
$|{\bf J}|$ calculated from the particle moments. We find no correlation between $\epsilon$ and $|{\bf J}|$, 
which suggests that regions of large $|{\bf J}|$, such as narrow current sheets, do not significantly enhance $\epsilon$ above background values. 
Similarly, we do not find any correlation between $\epsilon$ and ${\bf E} \cdot {\bf J}$ (not shown). 
Overall, aside from $\epsilon$ scaling with $T_e$ due to the correlation between $T_e$ and $n_e$, there is little statistical dependency of $\epsilon$ on the local plasma conditions. 

\begin{figure*}[htbp!]
\begin{center}
\includegraphics[width=160mm, height=80mm]{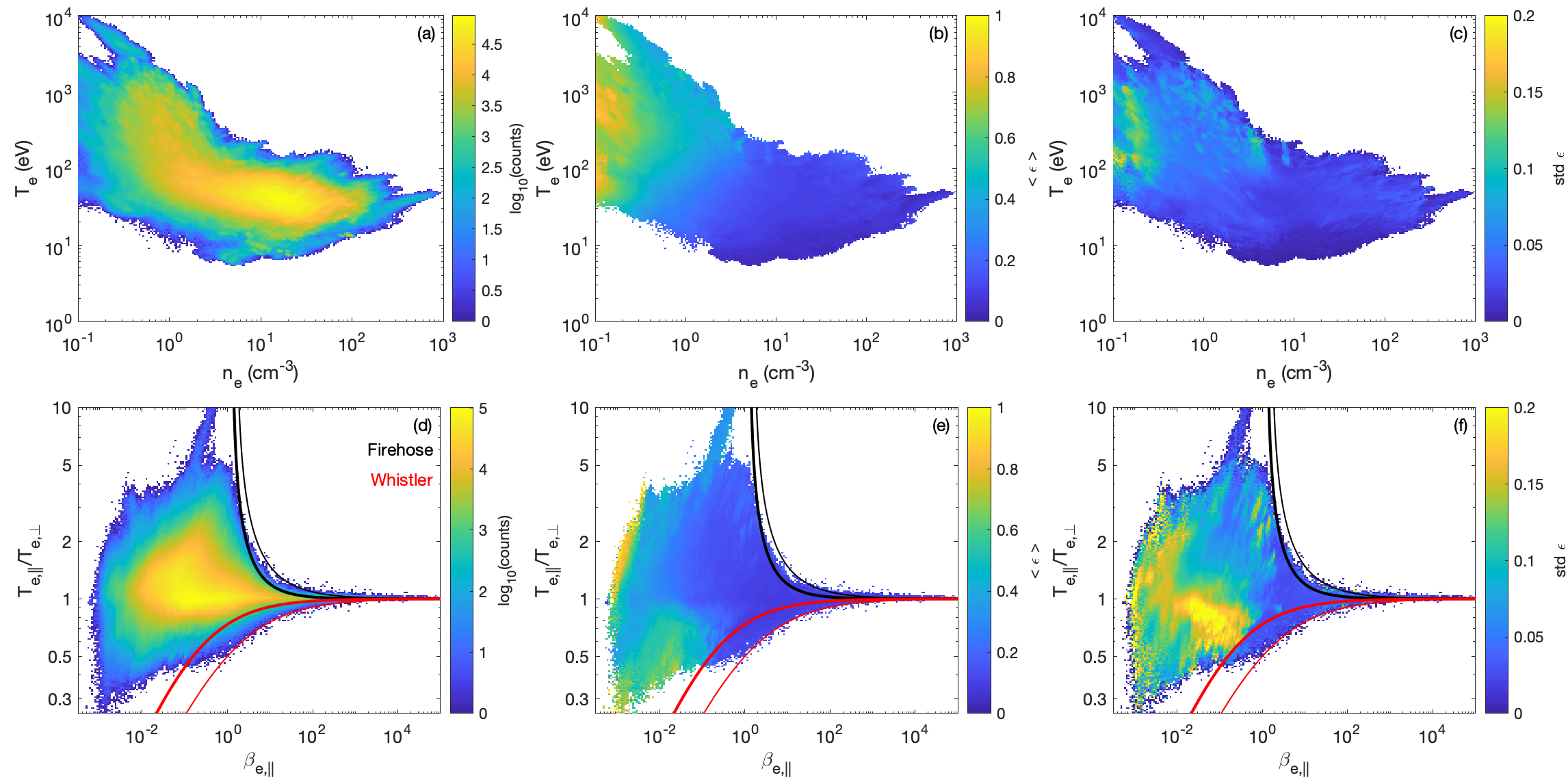}
\caption{Two-dimensional plots of counts and $\epsilon$ versus $T_e$ and $n_e$, and 
$T_{e,\parallel}/T_{e,\perp}$ and $\epsilon$. (a) Two-dimensional histograms of $T_e$ versus $n_e$. (b) Mean $\epsilon$ versus $T_e$ and $n_e$. 
(c) Standard deviation of $\epsilon$ versus $T_e$ and $n_e$. 
(d) Two-dimensional histogram of $T_{e,\parallel}/T_{e,\perp}$ versus $\beta_{e,\parallel}$. (e) 
Mean $\epsilon$ versus $T_{e,\parallel}/T_{e,\perp}$ versus $\beta_{e,\parallel}$. (f) 
Standard deviation of $\epsilon$ versus $T_{e,\parallel}/T_{e,\perp}$ versus $\beta_{e,\parallel}$.}
\label{statsbeta}
\end{center}
\end{figure*}

To further investigate the relationship between $\epsilon$ and $n_e$ and $T_e$ we plot the 2D histogram of 
$n_e$ and $T_e$ in Figure \ref{statsbeta}a. We see that there is a statistical increase in $T_e$ as $n_e$ decreases, as expected from electron distributions in the outer magnetosphere, magnetopause, and in the magnetosheath. 
However, for a given $n_e$ there is a large range of $T_e$, in particular for $n_e \lesssim 1$, corresponding to the outer magnetosphere. This larger range of $T_e$ in the magnetosphere can result from the relative densities of cold and hot components $n_c$ and $n_h$ varying significantly. 
In Figures \ref{statsbeta}b and \ref{statsbeta}c we plot the mean and standard deviation of $\epsilon$ versus 
$n_e$ and $T_e$. These values are computed on 
the same grid as in Figure \ref{statsbeta}a, so Figure \ref{statsbeta}a indicates the number of values used 
to calculate each mean and standard deviation. Figure \ref{statsbeta}b shows the strong dependence of 
$\epsilon$ on $n_e$, as well as a weaker dependence on $T_e$. In particular, the mean $\epsilon$ increases 
as $T_e$ increases for a given $n_e$. This is likely in part due to the lowest energy part of the electron distribution being excluded to avoid contamination from internal photoelectrons and spacecraft charging effects. Figure \ref{statsbeta}c shows that the standard deviation of $\epsilon$ tends to increase with decreasing $n_e$ and increasing $T_e$. This might correspond to more variable electron distributions in the outer magnetosphere. 

We now investigate the relationship between the parallel electron plasma $\beta$, $\beta_{e,\parallel}$, and 
$T_{e,\parallel}/T_{e,\perp}$ and the instability of these electron distributions. 
In Figure \ref{statsbeta}d we plot the 2D histogram of $T_{e,\parallel}/T_{e,\perp}$ and $\beta_{e,\parallel}$. 
We find that for $\beta_{e,\parallel} \lesssim 1$ there is a wide range of $T_{e,\parallel}/T_{e,\perp}$, 
although most data are found near $T_{e,\parallel}/T_{e,\perp} = 1$. For $\beta_{e,\parallel} \gtrsim 1$ the range 
of $T_{e,\parallel}/T_{e,\perp}$ decreases as $\beta_{e,\parallel}$ increases. 

We can compare this histogram with the thresholds for the oblique resonant electron firehose instability 
\cite[]{li3} and the whistler temperature anisotropy instability \cite[]{kennel1}. 
The oblique electron firehose instability can occur 
in a high $\beta_{e,\parallel}$ plasma with $T_{e,\parallel}/T_{e,\perp} > 1$. 
Theoretical work has shown that the oblique resonant electron firehose instability has a lower threshold 
than the field-aligned non-resonant electron fireshose instability \cite[]{li3}, 
so we only consider the former case here.
The whistler temperature anisotropy instability can occur 
for $T_{e,\parallel}/T_{e,\perp} < 1$. These instabilities are proposed to constrain the values of 
$T_{e,\parallel}/T_{e,\perp}$ that can develop by pitch-angle scattering electrons so the electron distribution becomes more isotropic. Numerical studies have shown that the threshold 
$T_{e,\parallel}/T_{e,\perp}$ for these instabilities scales with $\beta_{e,\parallel}$ and has the form
\cite[]{gary6}
\begin{equation}
\frac{T_{e,\parallel}}{T_{e,\perp}} = \left( 1 + S_e \beta_{e,\parallel}^{-\alpha_e} \right)^{-1},
\label{tempthres}
\end{equation}
where $S_e$ and $\alpha_e$ are constants. We consider the thresholds corresponding to growth rates 
$\gamma/\Omega_{ce} = 0.01$ and $0.1$. For the electron firehose instability we use $S_e = -1.23$ and 
$\alpha_e = 0.88$, and $S_e = -1.32$ and $\alpha_e = 0.61$ for $\gamma/\Omega_{ce} = 0.01$ and $0.1$, 
respectively \cite[]{gary5}. For the whistler instability we use $S_e = 0.36$ and $\alpha_e = 0.55$, and 
$S_e = 1.0$ and $\alpha_e = 0.49$ for $\gamma/\Omega_{ce} = 0.01$ and $0.1$, 
respectively \cite[]{gary6}. These thresholds are plotted in Figures \ref{statsbeta}d--\ref{statsbeta}f, 
where the black and red curves correspond to the firehose and whistler thresholds, respectively. The 
thick lines correspond to $\gamma/\Omega_{ce} = 0.01$ and the thin lines correspond to 
$\gamma/\Omega_{ce} = 0.1$. 

For $T_{e,\parallel}/T_{e,\perp} > 1$ the firehose instability for $\gamma/\Omega_{ce} = 0.01$ provides an 
approximate boundary for the largest $T_{e,\parallel}/T_{e,\perp}$ for $\beta_{e,\parallel} \gtrsim 2$. 
Only $0.016 \, \%$ of the data exceed the $\gamma/\Omega_{ce} = 0.01$ threshold, so regions unstable to 
the firehose instability are very rare. 
For $\beta_{e,\parallel} \lesssim 2$ the threshold for the firehose instability is much larger than any observed 
$T_{e,\parallel}/T_{e,\perp}$. This means that for $\beta_{e,\parallel} \lesssim 2$, such as on the 
magnetospheric side of the magnetopause and in the magnetosphere, the firehose instability is unlikely 
to occur, and thus cannot limit the values of $T_{e,\parallel}/T_{e,\perp}$ found there. Regions where 
the firehose instability is more likely to occur are in the magnetosheath, and potentially at the magnetopause 
boundary and in reconnection regions, where the magnetic field ${\bf B}$ becomes very small. 

For $T_{e,\parallel}/T_{e,\perp} < 1$ the whistler temperature anisotropy instability provides an approximate boundary for $\beta_{e,\parallel} \gtrsim 0.1$ for the allowable $T_{e,\parallel}/T_{e,\perp}$. 
For the $\gamma/\Omega_{ce} = 0.01$ threshold approximately $0.12 \, \%$ of the data exceed the threshold, 
nearly an order of magnitude higher than for the firehose instability. 
We find that for $\beta_{e,\parallel} \gtrsim 1$ the cutoff in the observed data matches the 
$\gamma/\Omega_{ce} = 0.1$ threshold very well. This might suggest that the maximum growth rate for whistlers near the magnetopause and in the magnetosheath is $\gamma/\Omega_{ce} = 0.1$. 

We note that the thresholds for the whistler and firehose instabilities are based on an electron bi-Maxwellian 
distribution function. Thus, a large deviation of the observed distribution from a bi-Maxwellian distribution may 
invalidate the threshold predictions. In Figures \ref{statsbeta}e and \ref{statsbeta}f 
we plot the mean and standard deviation of 
$\epsilon$ as functions of $T_{e,\parallel}/T_{e,\perp}$ and $\beta_{e,\parallel}$. 
Figure \ref{statsbeta}e shows that when $T_{e,\parallel}/T_{e,\perp}$ approaches or exceeds the whistler 
and firehose thresholds the values of $\epsilon$ remain relatively small. This suggests that the thresholds 
are reasonable indicators of instability. 
Figure \ref{statsbeta}f shows that the standard deviation is relatively 
small when the thresholds are exceeded, although there are regions with enhanced standard deviations close 
to both the firehose and whistler instabilities. This might indicate that some distributions in these regions 
may be unstable. 

For magnetosheath plasma, electron distributions often exhibit flat-top distribution functions so they will often differ 
from a bi-Maxwellian, producing a finite $\epsilon$. It is unclear to what to what extent such distributions affect the firehose and whistler thresholds, 
although the good agreement between the threshold conditions and the cutoff 
in the observed data suggests that the numerical thresholds for the instabilities are reasonable. 
For magnetospheric plasmas previous observations have shown that non-Maxwellian distributions develop. 
One example is the magnetospheric plasmas consist of both hot and cold electron distributions, which will modify the threshold for whistler waves \cite[e.g.][]{gary2}. Similarly, in the magnetospheric separatrices of reconnection complex electron distributions develop, which can excite whistler waves \cite[]{graham3,khotyaintsev2019}. 
Thus, we expect that 
whistlers can be generated if they do not exceed the threshold conditions due to the deviations from the 
bi-Maxwellian distribution function. It is unclear if deviations from a bi-Maxwellian distributions can enhance the instability of the oblique electron firehose instability. 

\section{Case studies} \label{casestudies}
Using the results in Figure \ref{statsne}, we can use the statistical percentiles of $\epsilon$ as a function of 
$n_e$ to find regions of localized enhanced or reduced $\epsilon$ and investigate their source regions. 
We specifically focus on the magnetic reconnection ion and electron diffusion regions, the bowshock, and the turbulent magnetosheath. 

\subsection{Ion Diffusion Region}
\begin{figure*}[htbp!]
\begin{center}
\includegraphics[width=160mm, height=150mm]{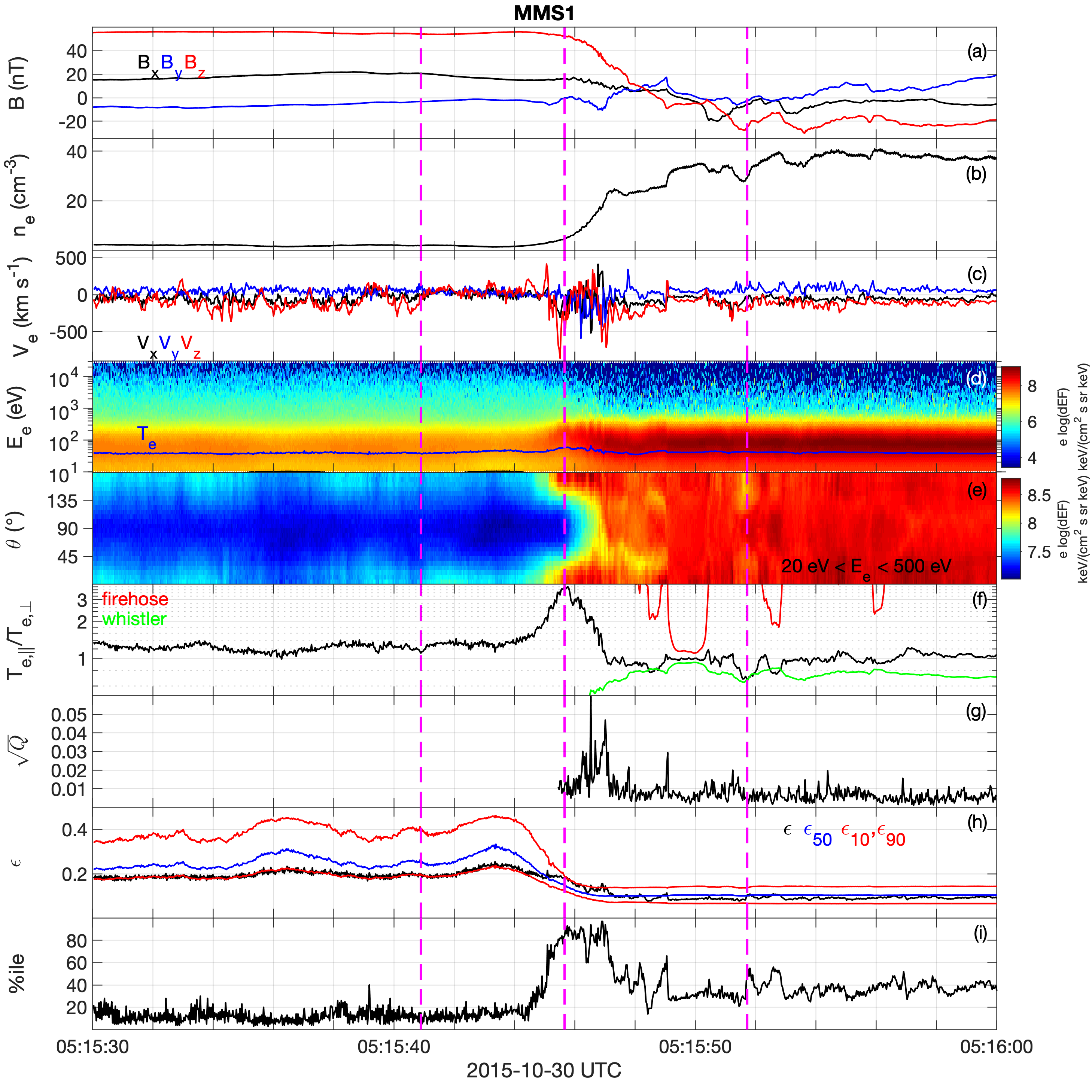}
\caption{Overview of the electron behavior from MMS1 associated with the ion diffusion region observed on 
2015 October 30. (a) ${\bf B}$. (b) $n_e$. (c) ${\bf V}_e$. (
d) Electron omnidirectional differential energy flux (blue line indicates $T_e$).  
(e) Spectrogram of the electron pitch-angle distribution for energies 
$20 \, \mathrm{eV} < E < 500 \, \mathrm{eV}$. (f) $T_{e,\parallel}/T_{e,\perp}$ (black), 
and firehose (red) and whistler (green) thresholds (for $\gamma/\Omega_{ce} = 0.01$). 
(g) $\sqrt{Q}$. (h) $\epsilon$ (black), and the 10th and 90th percentiles of $\epsilon$ as a function of 
$n_e$ (red) and median $\epsilon$ as a function $n_e$. (i) Percentile of the observed $\epsilon$ as a function 
of $n_e$. The magenta dashed lines indicate the times the electron distributions in 
Figure \ref{exampledist} are taken. }
\label{IDReg}
\end{center}
\end{figure*}

We investigate the ion diffusion region (IDR) observed 2015 December 30 by \cite{graham4} at Earth's dayside 
magnetopause. The IDR was identified by a strong Hall electric field and intense parallel electron heating. 
Figure \ref{IDReg} 
provides an overview of the electron behavior for this event from MMS1. The spacecraft crossed the 
magnetopause from the magnetosphere to the magnetosheath, indicated by the reversal in the magnetic field 
${\bf B}$ and substantial increase in $n_e$ (Figures \ref{IDReg}a and \ref{IDReg}b). 
At the magnetopause, we see large fluctuations in the electron bulk velocity ${\bf V}_e$ (Figure \ref{IDReg}c), 
intense parallel electron heating (Figure \ref{IDReg}f), and an increase in $\sqrt{Q}$ (Figure \ref{IDReg}g). 
We find that $\sqrt{Q}$ peaks at $0.06$ near the center of the current sheet, 
which may indicate close proximity to the EDR. 
The peak in electron heating was observed on the magnetospheric side of the magnetopause in the 
magnetospheric inflow region next to the Hall region, where ions decouple from the electron motion 
\cite[]{graham4}. Here $\beta_{e,\parallel}$ is low so the peak $T_{e,\parallel}/T_{e,\perp} = 3.9$ is well 
below the threshold for the electron firehose instability. Therefore, the electron firehose instability cannot limit the 
magnitude of $T_{e,\parallel}/T_{e,\perp}$ found in the magnetopause current sheet. On the magnetosheath side 
of the current sheet we find that some short intervals have $T_{e,\parallel}/T_{e,\perp} < 1$ that exceed 
the whistler temperature anisotropy instability, and find that there are whistler waves in these regions (not shown). 

The electron non-Maxwellianity $\epsilon$ is plotted in Figure \ref{IDReg}h. Overall, we find that $\epsilon$ 
decreases from about $0.2$ to about $0.1$ from the magnetosphere to the magnetosheath. This is due 
primarily to the change in $n_e$, so it is difficult to identity regions of enhanced non-Maxwellianity by simply 
plotting $\epsilon$. 
Using the measured $n_e$ we calculate the 10th, 
50th (median), and 90th percentiles of $\epsilon$, 
$\epsilon_{10}$, $\epsilon_{50}$, and $\epsilon_{90}$, respectively, from the statistical results in 
Figure \ref{statsne}c. These quantities are 
plotted in Figure \ref{IDReg}h. 
In the magnetosphere (low density) we find that $\epsilon$ remains close to $\epsilon_{10}$, 
indicating that statistically the distribution is close to Maxwellian for a magnetospheric distribution. 
In this case there is only a single colder electron population with $T_e \approx 40$~eV, and negligible hot electrons. This results in the relatively low $\epsilon$ observed in the magnetosphere. 
In the magnetosheath (high density) $\epsilon$ has 
values between $\epsilon_{10}$ and $\epsilon_{50}$, suggesting that while the electron distributions are non-Maxwellian, but the non-Maxwellianity is not particularly large. However, in the IDR where large $T_{e,\parallel}/T_{e,\perp}$ occurs there is an increase in $\sqrt{Q}$, and $\epsilon$ approaches and exceeds $\epsilon_{90}$, meaning that statistically the non-Maxwellianity is high in this region. This is most clearly seen in Figure \ref{IDReg}i, where we plot the percentile 
that the observed $\epsilon$ belongs to as a function of $n_e$ based on the statistical results in Figure \ref{statsne}c. We find that the percentile increases where $T_{e,\parallel}/T_{e,\perp}$ and $\sqrt{Q}$ peak, while in the magnetosphere the percentile is low and in the magnetosheath the percentile remains close to $50$. 
The magenta dashed lines in Figure \ref{IDReg} indicate the times of the three electron distributions in Figure \ref{exampledist}. 
The source of the increased $\epsilon$ in the ion diffusion region is the flat-top shape of the distribution parallel and antiparallel 
to ${\bf B}$ (Figure \ref{exampledist}d). Figure \ref{exampledist} shows that the shapes of the magnetospheric and magnetosheath distributions in the thermal energy range deviate from a bi-Maxwellian distribution. 

These results indicate that calculating $\epsilon$ alone is not sufficient to identify regions of unusually large non-Maxwellianity when considering magnetopause crossings. However, using the statistical results derived from Figure \ref{statsne}c, we can identify regions where 
the calculated $\epsilon$ correspond to unusually large deviations from a bi-Maxwellian distribution function. 
For this example enhanced deviations in the electron distributions from a bi-Maxwellian are found in the ion 
diffusion region of asymmetric reconnection, although there is little change in the value of $\epsilon$ across the magnetopause. 

\subsection{Electron Diffusion Region}
\begin{figure*}[htbp!]
\begin{center}
\includegraphics[width=160mm, height=150mm]{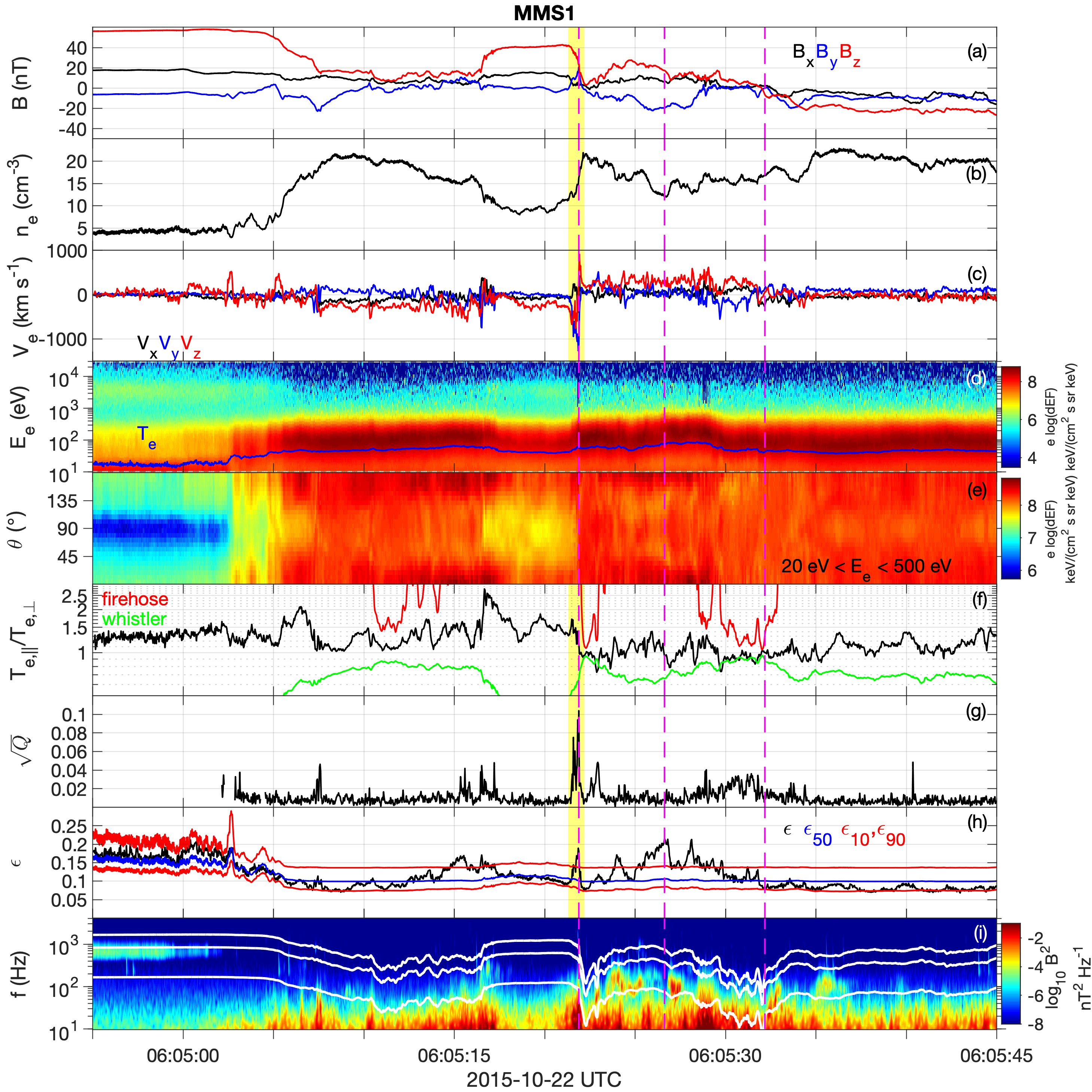}
\caption{Overview of the electron behavior from MMS1 associated with the electron diffusion region observed on 
2015 October 22. (a) ${\bf B}$. (b) $n_e$. (c) ${\bf V}_e$. (d) Electron omnidirectional differential energy flux (blue line indicates $T_e$).  
(e) Spectrogram of the electron pitch-angle distribution for energies 
$20 \, \mathrm{eV} < E < 500 \, \mathrm{eV}$. (f) $T_{e,\parallel}/T_{e,\perp}$ (black), 
and firehose (red) and whistler (green) thresholds (for $\gamma/\Omega_{ce} = 0.01$). 
(g) $\sqrt{Q}$. (h) $\epsilon$ (black), and the 10th and 90th percentiles of $\epsilon$ as a function of $n_e$ (red) and median $\epsilon$ as a function $n_e$. (h) Percentile of the observed $\epsilon$ as a function of 
$n_e$. }
\label{EDReg}
\end{center}
\end{figure*}

As an exampe of an EDR crossing, Figure \ref{EDReg} shows the magnetopause crossings observed on 2015 October 22 by MMS1 \cite[]{phan3,toledoredondo2}. 
The EDR is observed at 06:05:22 UT, indicated by the yellow-shaded region in Figure \ref{EDReg}. Over the entire interval we observe three partial magnetopause crossings, where $B_z$ decreases and $n_e$ increases 
(Figures \ref{EDReg}a and \ref{EDReg}b). Between the first and second magnetopause crossings there is a 
southward ion outflow and after the third magnetopause crossing, where the EDR is observed, 
there is a northward outflow. The spacecraft enters the magnetosheath at approximately 06:05:35 UT. 
The outflows are here indicated by the large-scale changes in the electron flow $V_{e,z}$ in Figure \ref{EDReg}c. 
In the magnetosphere $T_e$ is very low due to cold electrons dominating the electron distributions 
(Figure \ref{EDReg}d). The electron pitch-angle distribution in Figure \ref{EDReg}e shows that there are complex 
changes in the electron distribution functions across the magnetopause boundaries. 
Similarily, $T_{e,\parallel}/T_{e,\perp}$ varies significantly 
across the interval (Figure \ref{EDReg}f). 
At the third magnetopause crossing we observe the largest ${\bf V}_e$ (Figure \ref{EDReg}c) and $\sqrt{Q}$ (Figure \ref{EDReg}g), which 
peaks just above $0.1$, signifying the EDR. 

In Figure \ref{EDReg}h we plot the timeseries of $\epsilon$, as well as $\epsilon_{10}$, $\epsilon_{50}$, and 
$\epsilon_{90}$. We note that after the first magnetopause crossing $\epsilon_{10}$, $\epsilon_{50}$, and 
$\epsilon_{90}$ remain relatively constant. 
In the EDR we find a large enhancement in $\epsilon$ (well above $\epsilon_{90}$), which is colocated 
with the peak in $\sqrt{Q}$. For this event we also observe very large enhancements in $\epsilon$ in both 
the northward and southward ion outflows (both have extended regions where $\epsilon > \epsilon_{90}$). Indeed the largest 
$\epsilon$ is observed in the northward reconnection outflow, rather than at the EDR. Here, $\sqrt{Q}$ is 
negligible so these deviations from bi-Maxwellianity are gyrotropic in nature. This event illustrates why 
there is statistically a lack of correlation between $\epsilon$ and $\sqrt{Q}$. 

In Figure \ref{EDReg}f we compare $T_{e,\parallel}/T_{e,\perp}$ with the local electron firehose and whistler thresholds. 
Throughout the interval $T_{e,\parallel}/T_{e,\perp}$ remains below the threshold of the electron firehose instability. For the whistler instability we find 
that in some regions $T_{e,\parallel}/T_{e,\perp}$ exceeds the threshold. Figure \ref{EDReg}i, which 
shows the frequency-time spectrogram of ${\bf B}$, showing that whistler waves are found near these regions. 
However, whistlers are found where $T_{e,\parallel}/T_{e,\perp}$ does not satisfy the threshold for instability, such 
as in the magnetosphere and throughout the northward outflow. In the magnetosphere we observe distinct 
hot and cold electron populations (Figure \ref{EDReg}d), where the cold population is characterized by $T_{e,\parallel}/T_{e,\perp} > 1$, 
resulting in the overall distribution having $T_{e,\parallel}/T_{e,\perp} > 1$, while the hot population with energies $E > 1$~keV has $T_{e,\parallel}/T_{e,\perp} < 1$, which is the likely source of the whistler waves. 
In this case we estimate $n_h/n_c \approx 3 \times 10^{-3}$, meaning that the hot component has only a small 
effect on $\epsilon$. 
In the outflow we observe complex electron distributions (an example is shown below in Figure \ref{EDRedists}).  
Therefore, in these regions there tends to be a significant 
$\epsilon$, so the whistler instability thresholds predicted from a single bi-Maxweliian distribution are no longer valid. 

\begin{figure*}[htbp!]
\begin{center}
\includegraphics[width=160mm, height=140mm]{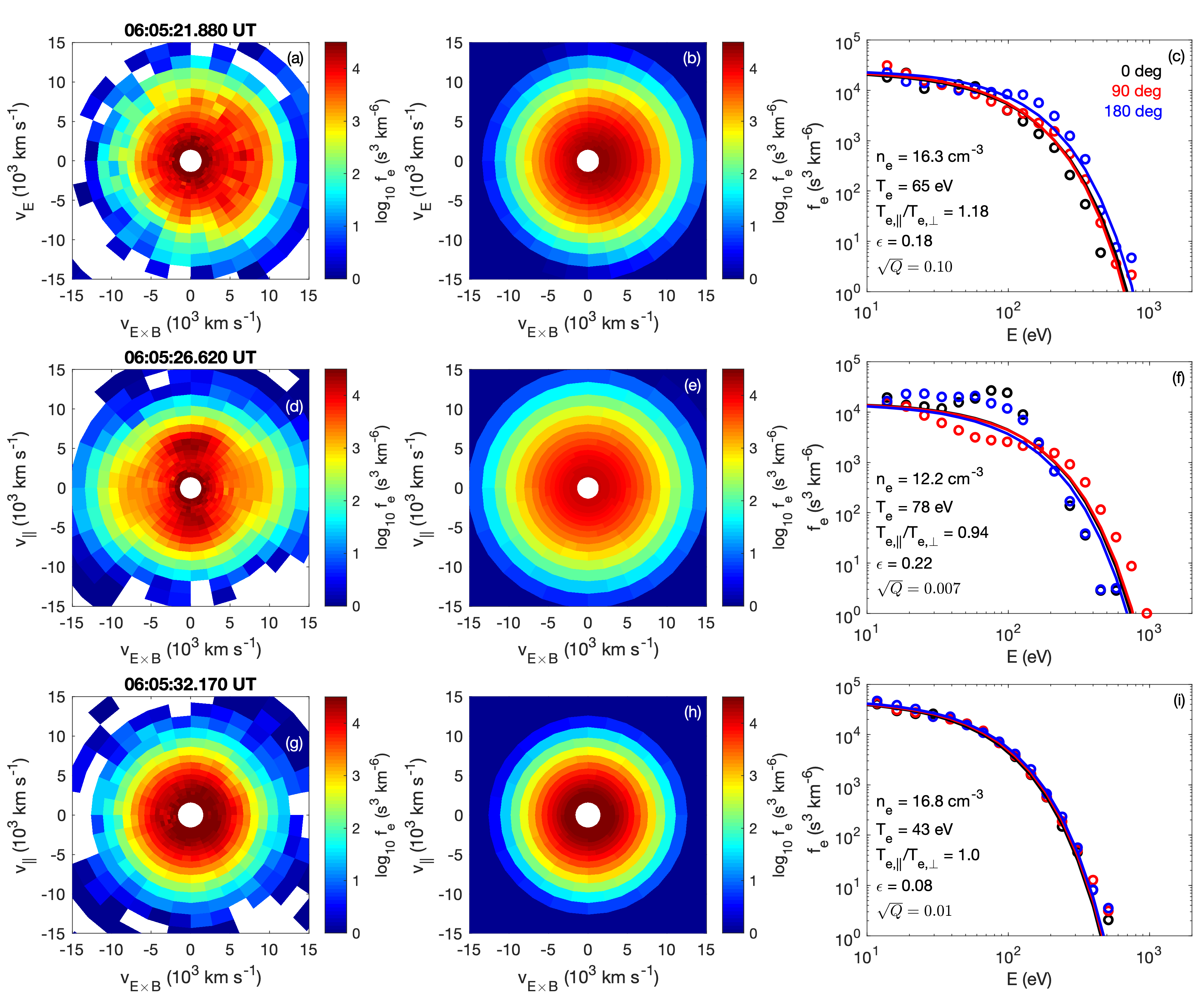}
\caption{Three examples of electron distributions and the predicted bi-Maxwellian distribution based on the electron moments from the EDR crossing observed by MMS1 on 22 October 2015. Panels (a), (d), and (g) show a two-dimensional slice of the observed three-dimensional electron distribution in the ${\bf B}$ and ${\bf E} \times {\bf B}$ plane. Panels (b), (e), and (h) show the modeled bi-Maxwellian distributions in the same plane. Panels (c), (f), and (i) show the phase-space densities at pitch angles $0^{\circ}$ (black), $90^{\circ}$ (red), and $180^{\circ}$ (blue) for the observed 
distributions (circles) and modeled bi-Maxwellian (solid lines). The distribution in panels (a)--(c) is in the EDR where $\sqrt{Q}$ peaks, (d)--(f) is in the ion outflow where $\epsilon$ peaks, (g)--(i) is close to the 
magnetosheath where $\epsilon$ is small. The electron distribution properties and $\epsilon$ 
of the three distributions are given in panels (c), (f), and (i).}
\label{EDRedists}
\end{center}
\end{figure*}

In Figure \ref{EDRedists} we investigate three electron distributions observed in the EDR where $\sqrt{Q}$ 
peaks (Figures \ref{EDRedists}a--\ref{EDRedists}c), in the reconnection outflow where $\epsilon$ peaks 
(Figures \ref{EDRedists}d--\ref{EDRedists}f), and a point in the ouflow where $\epsilon$ is small 
(Figures \ref{EDRedists}g--\ref{EDRedists}i). These points are indicated by the three dashed magenta lines 
in Figure \ref{EDReg}. Figure \ref{EDRedists}a shows a distribution consisting of a near-stationary core and a 
crescent propagating in the ${\bf E} \times {\bf B}$ direction, which is responsible for the peak in 
$\sqrt{Q}$ and the large ${\bf V}_e$. Such distributions are typical of EDRs at the magnetopause 
\cite[]{burch2,graham11,norgren4}. 
As expected the model distribution does not capture the observed features, and is simply a bi-Maxwellian 
drifting in the ${\bf E} \times {\bf B}$ direction, which results in the enhanced $\epsilon$. 
Note that the pitch-angle distribution (Figure \ref{EDRedists}c) does not capture the agyrotropy, so in this plot 
the source of $\epsilon$ becomes unclear. 

In Figure \ref{EDRedists}d the observed distribution prominantly feaatures counter-streaming electron beams 
parallel and antiparallel to ${\bf B}$, and higher-energy electrons perpendicular to ${\bf B}$. 
For electron energies $E \lesssim 200 \, \mathrm{eV}$ the counter-streaming result in 
$T_{e,\parallel}/T_{e,\perp} > 1$, while for $E \gtrsim 200 \, \mathrm{eV}$ we find 
$T_{e,\parallel}/T_{e,\perp} < 1$. The total $T_{e,\parallel}/T_{e,\perp}$ is close to $1$, so the model 
distribution is approximately Maxwellian (Figure \ref{EDRedists}e). Figure \ref{EDRedists}f shows 
that $f_e$ at pitch angles $0^{\circ}$, $90^{\circ}$, and $180^{\circ}$ all differ significantly from 
$f_{\mathrm{model}}$ and as a result $\epsilon$ is relatively large. These distributions can account for the observed whistler waves, where $T_{e,\parallel}/T_{e,\perp}$ does not satisfy the instability threshold. 

The distribution in Figure \ref{EDRedists}g is close to Maxwellian and very similar to $f_{\mathrm{model}}$ 
(Figure \ref{EDRedists}h). Figure \ref{EDRedists}i shows that the profiles of $f_e$ at $\theta = 0^{\circ}$, 
$90^{\circ}$, and $180^{\circ}$ match well $f_{\mathrm{model}}$, so the distribution is approximately 
Maxwellian. Figures \ref{EDReg} and \ref{EDRedists} show that the shape electron distributions can vary significantly 
throughout the reconnection outflow, with both complex distributions and approximately Maxwellian 
distributions developing. 

More generally we have investigated the non-Maxwellianity of the electron distributions associated with the 
11 EDRs observed in the first phase of the MMS mission \cite[]{fuselier2,webster1}. 
We find that in each case there is an enhancement in $\epsilon$ in or near the EDR, which typically exceeds $\epsilon_{90}$. 
In some outflow regions large $\epsilon$ are observed, although the values of $\epsilon$, and whether 
these values exceed $\epsilon_{90}$, varies between events. This suggests that large values of $\epsilon$ are a necessary, but not sufficient criterion for identification of EDRs. 

%In this subsection we investigate the non-Maxwellianity of electron distributions associated with electron diffusion 
%regions associated with magnetopause reconnection. For the data used in this study 11 EDRs were observed 
%\cite[]{fuselier2,webster1}. 

\subsection{Bowshock}
\begin{figure*}[htbp!]
\begin{center}
\includegraphics[width=160mm, height=150mm]{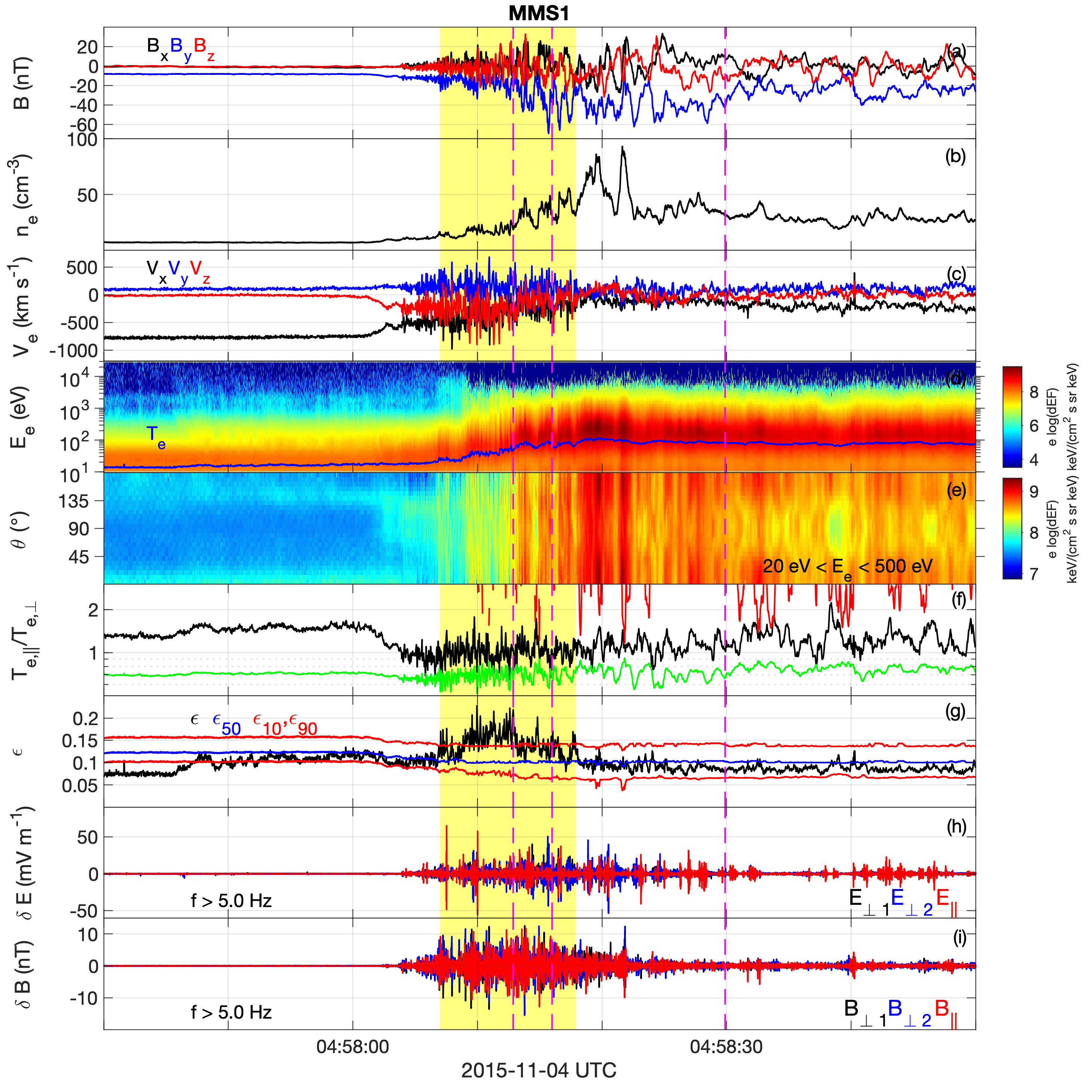}
\caption{Overview of the electron behavior from MMS1 associated a bowshock crossing observed on 
2015 November 4 observed by MMS1. (a) ${\bf B}$. (b) $n_e$. (c) ${\bf V}_e$. 
(d) Electron omnidirectional differential energy flux (blue line indicates $T_e$).  
(e) Spectrogram of the electron pitch-angle distribution for energies 
$20 \, \mathrm{eV} < E < 500 \, \mathrm{eV}$. (f) $T_{e,\parallel}/T_{e,\perp}$ (black), 
and firehose (red) and whistler (green) thresholds (for $\gamma/\Omega_{ce} = 0.01$). 
(g) $\epsilon$ (black), and the 10th and 90th percentiles of $\epsilon$ as a function of $n_e$ (red) and median $\epsilon$ as a function $n_e$. 
(h) Fluctuating electric field $\delta {\bf E}$ above $f = 5 \, \mathrm{Hz}$ in field-aligned coordinates. (i) Fluctuating magnetic field $\delta {\bf B}$ above $f = 5 \, \mathrm{Hz}$ in field-aligned coordinates. }
\label{Boweshockeg}
\end{center}
\end{figure*}

In this subsection we investigate the non-Maxwellianity of electron distributions at the bowshock. In particular, we investigate a quasi-perpendicular bowshock in detail. We 
find that $\epsilon$ is strongly enhanced at the bowshock but typically returns to nominal values within the 
magnetosheath.  
Figure \ref{Boweshockeg} shows an example of a quasi-perpendicular bowshock observed on 04 November 
2015 around 05:58:00 UT. This bowshock was previously studied by \cite{oka2017}. At this time the spacecraft 
were located at [10.4, 2.1, -0.5]~$R_E$ (GSE). For this shock we estimate the shock-normal angle to be 
$\theta_{Bn} = 83^{\circ}$, corresponding to a quasi-perpendicular shock, and the Alfven Mach number is $M_A 
\approx 11$. Based on the width of the shock foot we estimate that the shock moves $\sim 30$~km~s$^{-1}$ Sunward in 
the spacecraft frame. 

Figure \ref{Boweshockeg} shows that the spacecraft was initially in a solar wind-like plasma, with low ${\bf B}$,  
fast Earthward flow $V_{e,x} = -700$~km~s$^{-1}$, and density $n_e = 7.1$~cm$^{-3}$. We see that the 
electron fluxes vary near $100$~eV and observe large-amplitude emission of Langmuir waves at the local plasma frequency (not shown), indicating that the spacecraft is in the electron foreshock region. The shock foot begins at approximately 05.58:00 UT, and is seen as the increase in $n_e$ and $|B_y|$, and the decrease in $|V_{e,x}|$. 
The shock ramp begins at approximately 05:58:14 UT, and a series of shock ripples are observed in 
${\bf B}$ and $n_e$. The overshoot is observed at 05:58:19 UT.

In Figure \ref{Boweshockeg}g we plot $\epsilon$ along with $\epsilon_{10}$, $\epsilon_{50}$, and 
$\epsilon_{90}$. At the beginning of the interval $\epsilon$ is very low, but increases beginning at 05:57:46 UT. 
At this time $\epsilon$ likely varies because the spacecraft is in the electron foreshock, 
where electrons reflected and accelerated at the bowshock can cause $\epsilon$ to increase above the 
values in the unperturbed solar wind. In the shock foot $\epsilon$ begins to exceed $\epsilon_{90}$, which extends for about $13$ seconds until just after the shock ramp 
(yellow-shaded interval in Figure \ref{Boweshockeg}). Based on the shock speed this region of enhanced 
$\epsilon$ has a spatial scale of $\sim 400$~km, which corresponds to $2 \rho_i$ or $5~d_i$ based 
on the magnetosheath parameters, where $\rho_i$ and $d_i$ are the ion Larmor radius and ion inertial 
length, respectively. Thus, the region of enhanced $\epsilon$ extends over ion spatial scales.
Downstream of the shock $\epsilon$ typically remains between $\epsilon_{10}$ and $\epsilon_{50}$, with 
some local enhancements in $\epsilon$. Throughout this region we observe rapid fluctuations in
$T_{e,\parallel}/T_{e,\perp}$, although $T_{e,\parallel}/T_{e,\perp}$ does not exceed the threshold for 
the firehose instability and rarely satisfies the threshold for whistler waves. 

In Figures \ref{Boweshockeg}h and \ref{Boweshockeg}i we plot the fluctuating electric and magnetic fields, 
$\delta {\bf E}$ and $\delta {\bf B}$, in field-aligned coordinates above $5$~Hz. The region of enhanced $\epsilon$ coincides 
with the most intense $\delta {\bf E}$ and $\delta {\bf B}$. Fluctuations both parallel and perpendicular to 
the background magnetic field are observed. For $\delta {\bf E}$ the parallel fluctuations are typically 
observed near the ion plasma frequency $f_{pi}$, while the perpendicular fluctuations are observed at 
lower frequencies. The magnetic field fluctuations are primarily seen at low frequencies below the lower 
hybrid frequency. In the magnetosheath $\delta {\bf B}$ are significantly reduced, while $\delta {\bf E}$ 
parallel to the background magnetic field continue to be observed intermittently. 
The large-amplitude electrostatic and electromagnetic 
fluctuations in the bowshock may result in the electron distribution with large $\epsilon$ becoming more Maxwellian 
in the magnetosheath due to wave-particle interactions. 

\begin{figure*}[htbp!]
\begin{center}
\includegraphics[width=160mm, height=140mm]{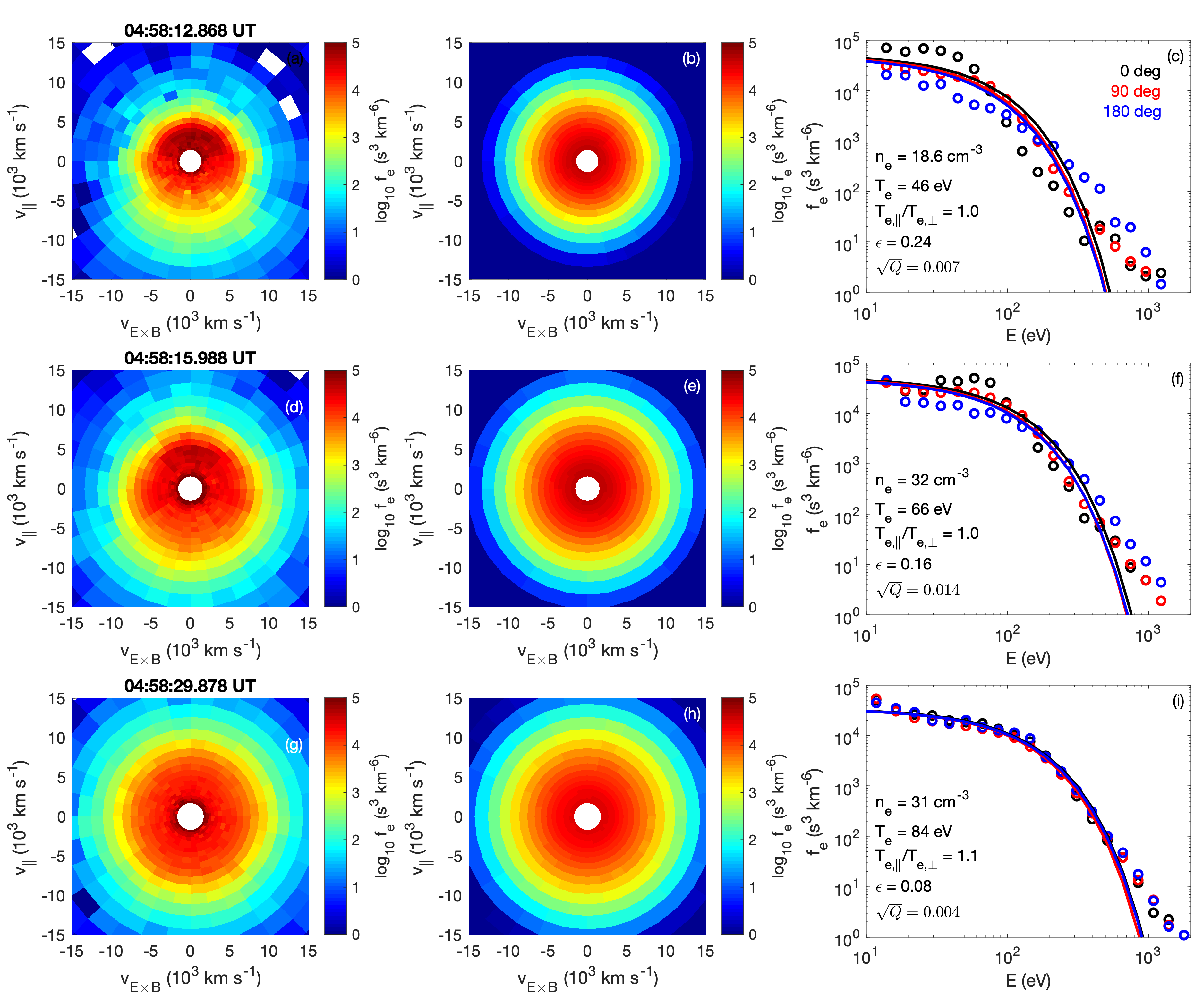}
\caption{Three examples of electron distributions and the modeled bi-Maxwellian distribution based on the electron moments from the foreshock crossing observed by MMS1 on 04 November 2015. Panels (a), (d), and (g) show a two-dimensional slice of the observed three-dimensional electron distribution in the ${\bf B}$ and ${\bf E} \times {\bf B}$ plane. Panels (b), (e), and (h) show the modeled bi-Maxwellian distributions in the same plane. Panels (c), (f), and (i) show the phase-space densities at pitch angles $0^{\circ}$ (black), $90^{\circ}$ (red), and $180^{\circ}$ (blue) for the observed 
distributions (circles) and modeled bi-Maxwellian (solid lines). The distribution in panels (a)--(c) is in the shock foot, (d)--(f) is near the shock ramp, (g)--(i) is in the magnetosheath behind the bowshock. 
The electron distribution properties and $\epsilon$ 
of the three distributions are given in panels (c), (f), and (i).}
\label{foreshockedists}
\end{center}
\end{figure*}

In Figure \ref{foreshockedists} we plot three electron distributions at the times indicated by the magenta 
dashed lines in Figure \ref{Boweshockeg} and compare these distributions with the modeled bi-Maxwellian distribution function. 
The distributions are observed in the shock foot [panels (a)--(c)], near the shock ramp [panels (d)--(f)], 
and in the magnetosheath [panels (g)--(i)]. 
In each observed distribution there is a super-thermal tail in the electron distribution. However, these tails 
do not significantly increase $\epsilon$ because the phase-space densities are very low, meaning their contributions to $T_e$ and $n_e$ are negligible. 

The electron distribution in Figure \ref{foreshockedists}a exhibits a dense low-energy electron beam parallel to ${\bf B}$, associated with accelerated solar wind electrons. For the direction antiparallel to ${\bf B}$ the electrons are hotter than in the parallel direction. These features are not captured by the bi-Maxwellian distribution (Figure \ref{foreshockedists}b), resulting in a relatively large $\epsilon$. 
Figure \ref{foreshockedists}c shows that the shape of the observed electron distribution differs greatly from 
the bi-Maxwellian distribution, consistent with $\epsilon$ being unusually large. 
Figure \ref{foreshockedists}d shows an electron distribution near the ramp, where shock ripples are observed. 
The distribution is similar to the one in Figure \ref{foreshockedists}a, except the solar wind electrons have been further accelerated parallel to ${\bf B}$ and now occupy a smaller range of pitch angles. 
At $\theta = 90^{\circ}$, and $180^{\circ}$ we observe flat-top like distributions, as previously observed 
at quasi-perpendicular shocks \cite[]{feldman1982,scudder1995}. 
Here, $\epsilon$ is reduced from the distribution in Figure \ref{foreshockedists}a, but remains well above the 
statistical median. These distributions consisting of a beam of accelerated solar wind electrons and flat-top-like 
distributions have been observed previously and are due to the cross-shock potential in the deHoffman-Teller frame. By integrating the divergence of the electron pressure divergence over position, we estimate a maximum cross-shock potential of $\sim 300$~V, which is several times larger than the maximum energy of the beam of 
solar wind electrons in the shock and the energies where the distribution is relatively flat $E \lesssim 100$~eV. 

In the magnetosheath behind the bowshock the electrons are close to Maxwellian, for example the electron distribution in Figures \ref{foreshockedists}g--\ref{foreshockedists}i. In these panels we see little deviation from a Maxwellian distribution function. We conclude that the strongly enhanced $\epsilon$ occur at ion spatial scales across the shock. Behind the shock in the magnetosheath $\epsilon$ is significantly reduced, but continues to vary with position.

\subsection{Turbulent Magnetosheath}
\begin{figure*}[htbp!]
\begin{center}
\includegraphics[width=160mm, height=150mm]{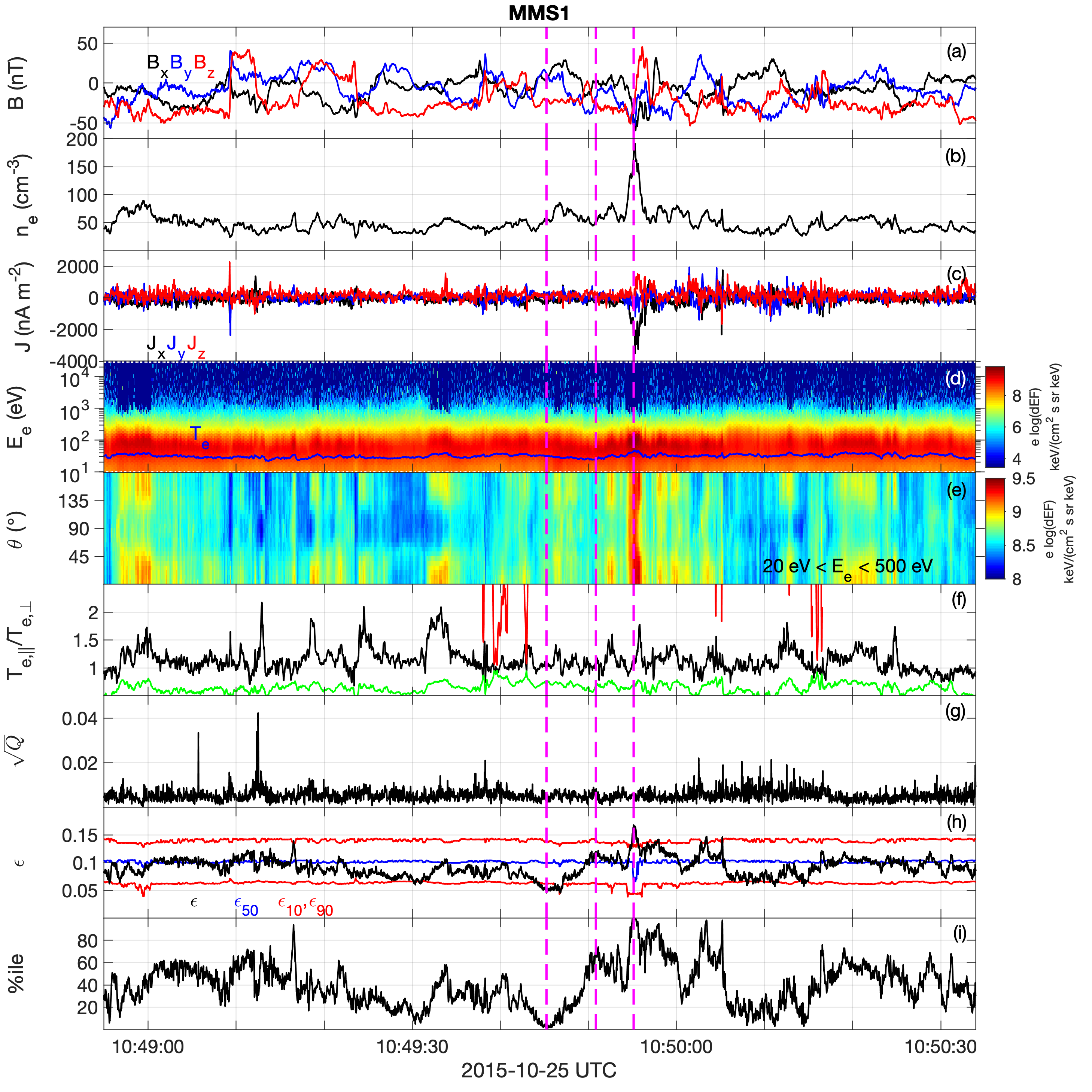}
\caption{Overview of the electron behavior from MMS1 in the turbulent magnetosheath observed on 
2015 October 25 observed by MMS1. (a) ${\bf B}$. (b) $n_e$. (c) ${\bf V}_e$. 
(d) Electron omnidirectional differential energy flux (blue line indicates $T_e$).  
(e) Spectrogram of the electron pitch-angle distribution for energies 
$20 \, \mathrm{eV} < E < 500 \, \mathrm{eV}$. (f) $T_{e,\parallel}/T_{e,\perp}$ (black), 
and firehose (red) and whistler (green) thresholds (for $\gamma/\Omega_{ce} = 0.01$). 
(h) $\epsilon$ (black), and the 10th and 90th percentiles of $\epsilon$ as a function of $n_e$ (red) and median $\epsilon$ as a function $n_e$. 
(i) Percentile of the observed $\epsilon$ as a function of $n_e$.}
\label{MSHturbeg}
\end{center}
\end{figure*}

In this subsection we investigate a region of magnetosheath turbulence behind the quasi-parallel bowshock. 
Figure \ref{MSHturbeg} provides an overview of the region observed by MMS1 on 25 October 2015. 
The interval is characterized by multiple current sheets, as indicated by reversal in ${\bf B}$ 
(Figure \ref{MSHturbeg}a), and narrow enhancements in ${\bf J}$ (Figure \ref{MSHturbeg}c). 
Figure \ref{MSHturbeg}d shows that $T_e$ remains relatively constant, although there are some variations 
in the electron fluxes. Figure \ref{MSHturbeg}e and \ref{MSHturbeg}f that $T_{\parallel}/T_{\perp}$ varies 
across the interval, with $T_{\parallel}/T_{\perp}$ ranging from $1$ to $2$. We find that the whistler and 
firehose thresholds for instability are not satisfied in this interval.
Throughout the interval $\sqrt{Q}$ remains relatively small, although some enhancements in $\sqrt{Q}$ 
occur over very short intervals. 

\begin{figure*}[htbp!]
\begin{center}
\includegraphics[width=120mm, height=100mm]{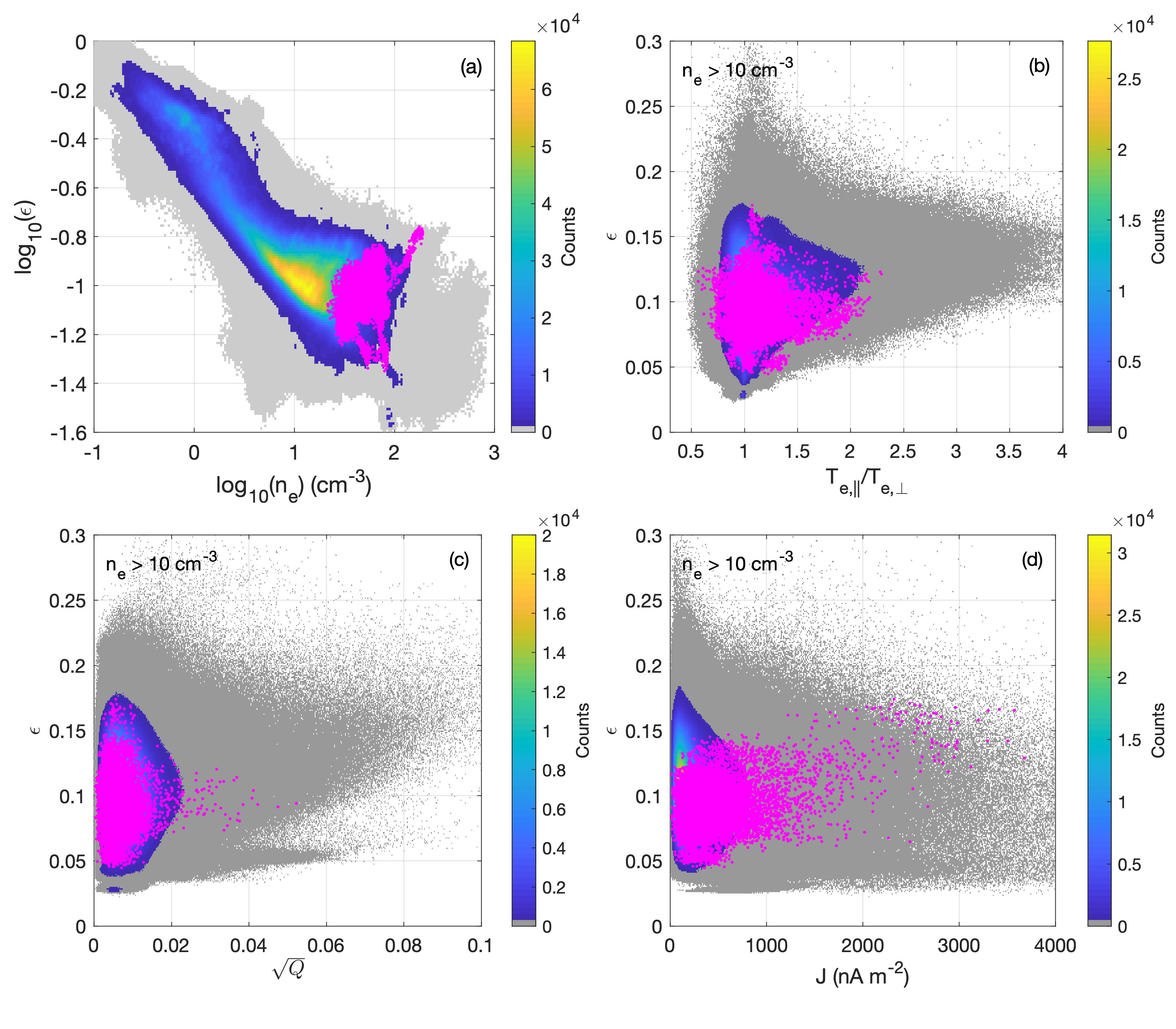}
\caption{Two-dimensional histograms of $\epsilon$ versus different plasma conditions when 
$n_e > 10$~cm$^{-3}$ [panels (b)--(d)]. Overplotted in magenta are scatter plots of the points in the interval shown in 
Figure \ref{Boweshockeg} for MMS1--MMS4. 
(a) $\log_{10}\epsilon$ versus $\log_{10}T_e$. (b) $\epsilon$ versus $\log_{10}T_{e,\parallel}/T_{e,\perp}$. 
(c) $\epsilon$ versus agyrotropy measure $\sqrt{Q}$. 
(d) $\epsilon$ versus $|{\bf J}|$.}
\label{statspropsMSH}
\end{center}
\end{figure*}

In Figure \ref{MSHturbeg}h we plot $\epsilon$, along with $\epsilon_{10}$, $\epsilon_{50}$, and $\epsilon_{90}$.  
In general, $\epsilon_{10}$, $\epsilon_{50}$, and $\epsilon_{90}$ remain relatively constant across the entire 
interval. We find that $\epsilon$ varies between $0.05$ and $0.17$, and peaks in the region of enhanced 
$n_e$ and ${\bf J}$ at 10:49:55 UT. Figure \ref{MSHturbeg}i shows that the percentile of $\epsilon$ ranges 
from $1$ to $99$, meaning the electron distributions vary from very close to Maxwellian to highly non-Maxwellian for magnetosheath conditions. 
In general, we see no clear correlation between $\epsilon$ and the other local plasma parameters. 
For example, there is no clear consistent correlation with ${\bf J}$. At 10:49:55 UT there is a peak in $\epsilon$ and ${\bf J}$; however, at other times there are peaks in ${\bf J}$ without enhanced $\epsilon$ (e.g., at 10:49:09 UT), and there are peaks in $\epsilon$ where ${\bf J}$ is negligible (e.g., at 10:49:16 UT). 
To illustrate this further, Figure \ref{statspropsMSH} shows the histograms of the entire dataset, with 
the data from the magnetosheath turbulence interval (from all four spacecraft) overplotted as magenta points. 
Figure \ref{statspropsMSH}a shows a wide range of $\epsilon$ observed in this interval and no clear correlation with $n_e$. Similarly, 
there is no correlation of $\epsilon$ with $T_{\parallel}/T_{\perp}$ and $\sqrt{Q}$ (Figures \ref{statspropsMSH}b and \ref{statspropsMSH}c). Figure \ref{statspropsMSH}d shows there is a slight tendency of $\epsilon$ to increase with ${\bf J}$, which in this case is due to the 
region centered around 10:49:55 UT. Overall, there is little clear correlation of $\epsilon$ with the local plasma conditions. Other turbulent magnetosheath intervals, such as those investigated in 
\cite{yordanova2016} and \cite{voros2017} yield qualitatively similar results. 
We propose that the enhanced non-Maxwellian features may develop at the bowshock, and that 
the locally observed changes in $\epsilon$ are due to the changing magnetic connectivity to the bowshock during the turbulent intervals. 

\begin{figure*}[htbp!]
\begin{center}
\includegraphics[width=160mm, height=140mm]{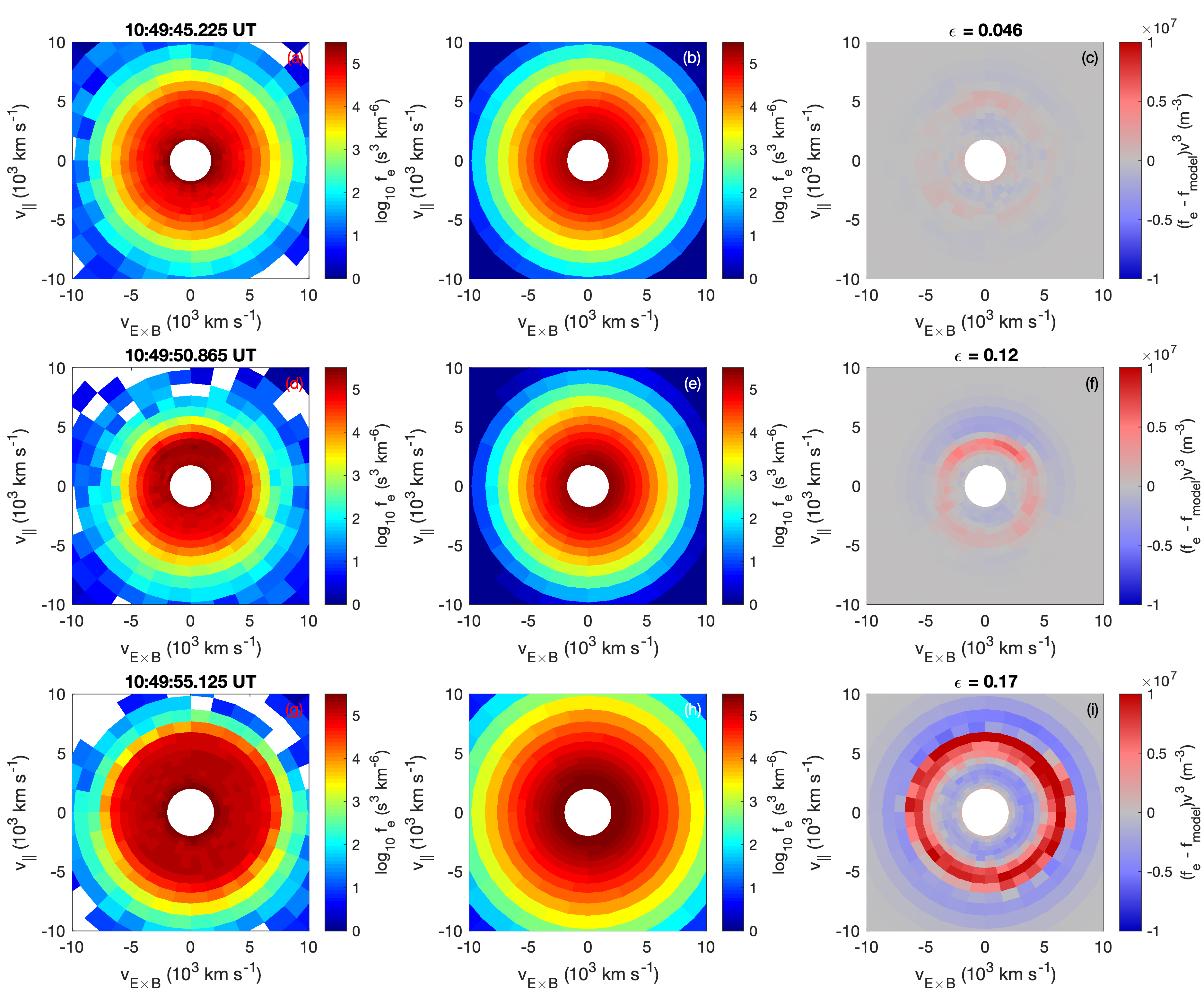}
\caption{Three examples of electron distributions and the predicted bi-Maxwellian distribution based on the electron moments from the foreshock crossing observed by MMS1 on 25 November 2015 in the turbulent magnetosheath. 
Panels (a), (d), and (g) show a two-dimensional slice of the observed three-dimensional electron distribution in the ${\bf B}$ and ${\bf E} \times {\bf B}$ plane. Panels (b), (e), and (h) show the modeled bi-Maxwellian distributions in the same plane. Panels (c), (f), and (i) show the modeled bi-Maxwellian distributions in the same plane. Panels (c) and (f) show 
$(f_e - f_{\mathrm{model}}) v^3$ in the same plane, which indicates the regions of velocity space that contribute most 
to $\epsilon$. The distribution in panels (a)--(c) is corresponds to low $\epsilon$, (d)--(f) correspond to moderate 
$\epsilon$, and (g)--(i) correspond to high $\epsilon$. 
The values of$\epsilon$ 
of the three distributions are given in the titles of panels (c), (f), and (i).}
\label{MSHedists}
\end{center}
\end{figure*}

In Figure \ref{MSHedists} we present three electron distributions at the times indicated by the magenta dashed 
lines in Figure \ref{MSHturbeg}. Each row consists of the observed electron distribution the model bi-Maxwellian 
distribution and $(f_e - f_{\mathrm{model}}) v^3$, which indicates the regions of velocity space that contribute most to $\epsilon$. The observed distributions have $\epsilon$ of $0.046$, $0.12$, and $0.17$, 
which correspond to percentiles of $1$, $72$, and $99$, respectively. In each case the distributions are close 
to isotropic, with $T_{\parallel}/T_{\perp} = 1.1$. For the first distribution, Figures \ref{MSHedists}a and 
\ref{MSHedists}b show that $f_e$ is very similar to $f_{\mathrm{model}}$, and as a result 
$(f_e - f_{\mathrm{model}}) v^3$ remains small. For the second there is a beam-like feature in $f_e$ (Figure \ref{MSHedists}d), which is not captured by $f_{\mathrm{model}}$ (Figure \ref{MSHedists}e) and results in an enhanced $\epsilon$. 
Figure \ref{MSHedists}g shows a clear flat-top distribution, which differs significantly from $f_{\mathrm{model}}$ 
(Figure \ref{MSHedists}h). 
Figure \ref{MSHedists}i shows large values of $(f_e - f_{\mathrm{model}}) v^3$, with 
$(f_e - f_{\mathrm{model}}) v^3 < 0$ at low speeds, $(f_e - f_{\mathrm{model}}) v^3 > 0$ at intermediate speeds, 
and $(f_e - f_{\mathrm{model}}) v^3 < 0$ at higher speeds. This results in a large $\epsilon$. This distribution is similar to the flat-top distributions found at the bowshock, which exhibit large $\epsilon$.

In summary, we find that in magnetosheath turbulence $\epsilon$ varies significantly, but is not directly correlated with local plasma conditions, such as density, temperature, temperature anisotropy, and current density. This may suggest that the local plasma turbulence does not significantly enhance $\epsilon$. We propose that the enhanced $\epsilon$ seen intermittently in magnetosheath turbulence is likely generated at the bowshock. We suggest that the changes in magnetic connectivity to the bowshock, due to the changes in the direction of ${\bf B}$ throughout magnetosheath turbulence affect the local values of $\epsilon$. 

%\section{Discussion}

\section{Conclusions} \label{conclusions}
In this paper we have investigated the deviation of electron distributions from the bi-Maxwellian distribution function. 
We have defined a dimensionless quantity $\epsilon$, which quantifies the deviation of the observed electron 
distribution from the bi-Maxwellian distribution function. We have calculated this quantity for the electron 
distributions observed by the four MMS over a six month interval, primarly focussing on the magnetosphere, magnetopause, and magnetosheath. The key results of this study are:

(1) The electron non-Maxwellianity $\epsilon$ scales inversely with the electron number density near the 
magnetopause. This is primarily due to the tendency of low-density magnetospheric electron distributions having distinct cold and hot populations, which deviate significantly from a single bi-Maxwellian distribution resulting in large $\epsilon$, 
whereas in the higher-density magnetosheath the electron distributionsa are characterized a single temperature.  By comparing specific events with the statistical study of $\epsilon$ versus number density we can identify regions of enhanced non-Maxwellianity.

(2) Statistically, the electron non-Maxwellianity does not depend strongly on local plasma properties such as temperature anisotropy, agyrotropy, and current density. 

(3) The observed temperature anisotropies are bounded by the thresholds for the oblique electron firehose instability and the whistler temperature anisotropy instability in high $\beta$ plasmas, such as at the magnetopause and in the magnetosheath. The distribitions close to these thresholds tend to be close to bi-Maxwellian. These results suggest that these instabilities constrain the electron temperature anisotropies that can develop in high-$\beta$ plasmas. 

(4) Enhanced $\epsilon$ are found in the ion and electron diffusion regions of magnetic reconnection. While very large $\epsilon$ are observed in the electron diffusion region where electron distributions are agyrotropic, similarly large $\epsilon$ can develop in the outflow regions of magnetic reconnection. Thus, $\epsilon$ cannot uniquely identify electron diffusion regions.

(5) Enhanced $\epsilon$ develops at the bowshock, due to the development of flat-top distributions and electron beams resulting from the cross-shock potential. These $\epsilon$ develop over ion spatial scales and tend to decrease behind the bowshock in the magnetosheath.

(6) Intermittent enhancements in $\epsilon$ are observed in magnetosheath turbulence. These increases in $\epsilon$ are not well correlated with the local plasma conditions, which might suggest that the observed $\epsilon$ are produced at the bowshock, and is highly variable in magnesheath turbulence due to the changing magnetic connectivity to the bowshock. 

These results show that $\epsilon$ can be used to identify regions where large deviations in the observed distributions from a bi-Maxwellian distribution function, which may suggest that local kinetic processes are occurring. 
Future work on electron non-Maxwellianity should include the following: 

(1) More detailed investigations of $\epsilon$ in magnetosheth turbulence and how it relates to local turbulent processes and connectivity to the bowshock. 

(2) Direct observation of the electron firehose instability and the resulting waves. Our results show that the 
electron temperature anisotropy is well constrained by the oblique electron firehose instability; however, to our 
knowledge the electron firehose instability has not been directly observed in spacecraft data. 

\acknowledgments
We thank the entire MMS team and instrument PIs for data access and support. 
This work was supported by 
the Swedish National Space Agency (SNSA), grant 128/17. 
MMS data are available at https://lasp.colorado.edu/mms/sdc/public.

\appendix \label{app1}
\section{Non-Maxwellianity of time-averaged electron distributions in the magnetosphere} \label{app1}
\begin{figure*}[htbp!]
\begin{center}
\includegraphics[width=120mm, height=100mm]{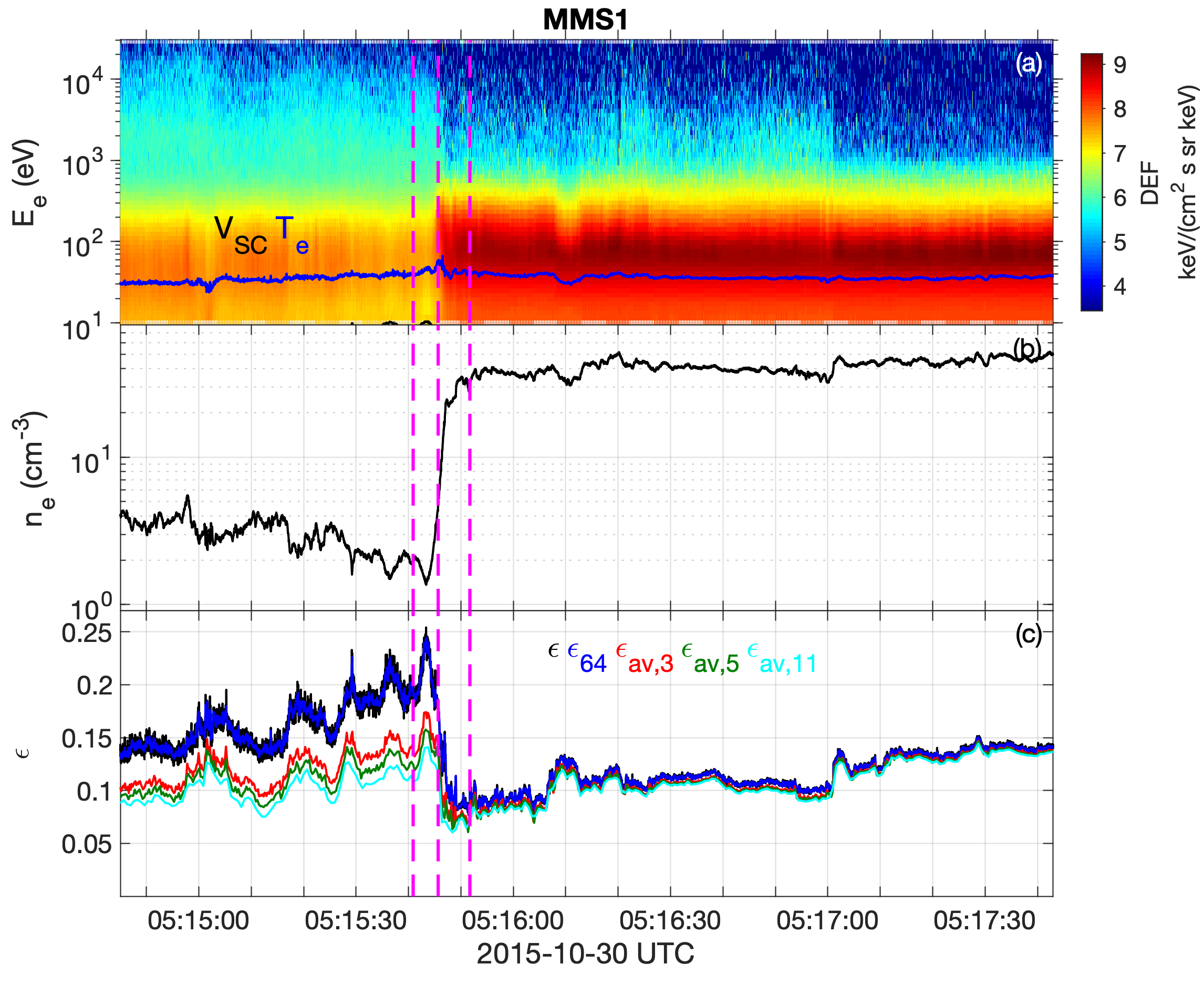}
\caption{Calculations of $\epsilon$ for different time averages for the magnetopause crossing observed on 2015 October 30 by MMS1. (a) Electron omnidirectional energy spectrogram with $T_e$ (blue) and $V_{SC}$ (black). (b) $n_e$. (c) Non-Maxwellianities $\epsilon$ (black), $\epsilon_{64}$, $\epsilon_{av,3}$ (red), $\epsilon_{av,5}$ (green), 
$\epsilon_{av,11}$ (cyan). The magenta dashed lines indicate the times of the electron distributions shown in Figure \ref{exampledist}. }
\label{epsav2015Oct30}
\end{center}
\end{figure*}

\begin{figure*}[htbp!]
\begin{center}
\includegraphics[width=120mm, height=100mm]{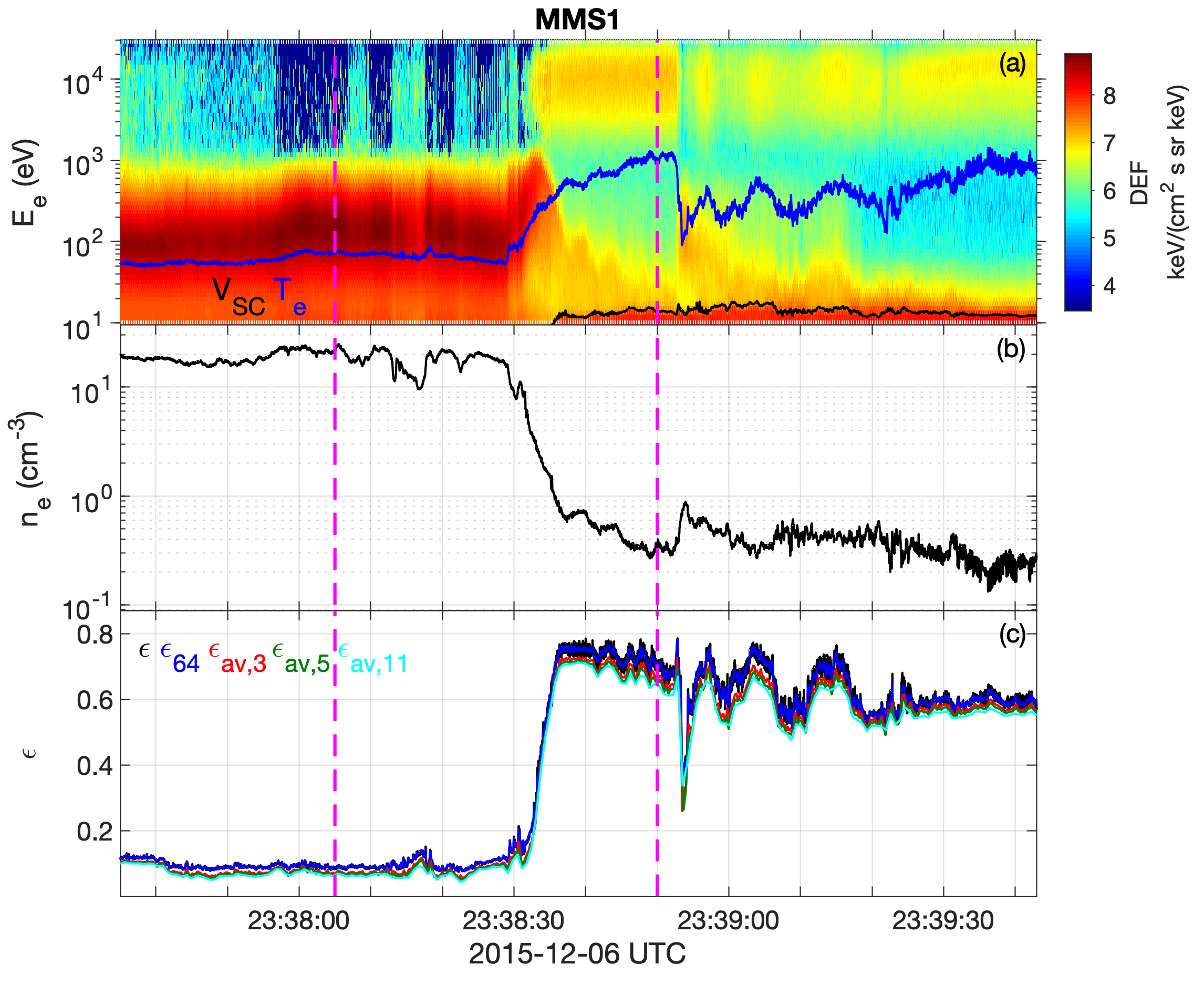}
\caption{Calculations of $\epsilon$ for different time averages for the magnetopause crossing observed on 2015 December 06 by MMS1. (a) Electron omnidirectional energy spectrogram with $T_e$ (blue) and $V_{SC}$ (black). (b) $n_e$. (c) Non-Maxwellianities $\epsilon$ (black), $\epsilon_{64}$, $\epsilon_{av,3}$ (red), $\epsilon_{av,5}$ (green), 
$\epsilon_{av,11}$ (cyan). The magenta dashed lines indicate the times of the electron distributions shown in Figure \ref{distneex}. }
\label{epsav2015Dec06}
\end{center}
\end{figure*}

In this Appendix we investigate the increase in $\epsilon$ in the magnetosphere due to the statistical noise in the electron distributions measured by FPI-DES. To do this we average the electron distributions over time to increase the overall counting statistics of the electron distributions and compare the results with the unaveraged distributions. The time averaging ensures that the observed distribution is smoother as a function of speed and angle. For phase 1a of the MMS mission FPI was operating in a mode where two distinct energy tables are used, which alternate for each consecutive electron distribution \cite[]{pollock1}. This results in the electrons being sampled at 64 energies over two consecutive electron distributions. To create time-averaged electron distributions we first combine each two consecutive electron distributions to construct electron 
distributions with 64 energy channels and sampled every $60$~ms. We then simply average $f_e$ of these distributions at all energies and angles to obtain the time-averaged electron distributions. 
We now calculate the non-Maxwellianity for the original distributions $\epsilon$, the distributions with 64 energy channels $\epsilon_{64}$, and electron distributions averaged over 3 $\epsilon_{av,3}$, 5 $\epsilon_{av,5}$, 
and 11 $\epsilon_{av,11}$ of the 64 energy channel distributions. Figures \ref{epsav2015Oct30} and 
\ref{epsav2015Dec06} show two examples of these calculations of $\epsilon$. 

In Figure \ref{epsav2015Oct30} we plot the magnetopause crossing shown in Figure \ref{IDReg}. Figures \ref{epsav2015Oct30}a and \ref{epsav2015Oct30}b show the electron energy spectrogram and $n_e$, respectively. The low-density magnetosphere is characterized by a single dominant electron distribution with temperature comparable to the magnetosheath. Figure \ref{epsav2015Oct30}c shows $\epsilon$, $\epsilon_{64}$, $\epsilon_{av,3}$, $\epsilon_{av,5}$, and $\epsilon_{av,11}$. We find that there is little difference between $\epsilon$ and 
$\epsilon_{64}$, except that the fluctuations in $\epsilon_{64}$ are smaller. This is not surprising because $\epsilon_{64}$ does not involve time averaging, so the counting 
statistics are not improved for the 64 energy channel distributions. However, we find that when the distributions are time-averaged there is a decrease in $\epsilon$ in the magnetosphere. In the magnetosheath, where $n_e$ is larger, time-averaging has only a very minor effect. In the magnetosphere the most significant decrease occurs between $\epsilon$ and $\epsilon_{av,3}$; taking longer time averages does not result in significant further decreases in $\epsilon$. In the magnetosphere $\epsilon$ decreases by $\sim 30$~\% when time-averaged distributions are used. This indicates that the counting statistics in the magnetosphere artificially increase $\epsilon$ in this case.

As a second example, we plot the magnetopause crossing observed on 2015 December 06 in Figure \ref{epsav2015Dec06}. Two electron distributions from this event are shown in Figure \ref{distneex}. For this event the magnetospheric electron distributions are composed of distinct hot and cold populations (Figure \ref{epsav2015Oct30}a), in contrast to the event in Figure \ref{epsav2015Oct30}. In this case time-averaging only results in a very small decrease in $\epsilon$ both in the magnetosheath and in the magnetosphere. In the magnetosphere the very large $\epsilon$ results from the electrons having distinct temperatures of $\sim 10$~eV 
and $\gtrsim 1$~keV. In this case the effect of counting statistics on $\epsilon$ is very small, and the observed $\epsilon$ is physical in the magnetosphere. In this case the distinct electron temperatures of magnetospheric 
electron distributions primarily determines $\epsilon$. 

\begin{figure*}[htbp!]
\begin{center}
\includegraphics[width=120mm, height=100mm]{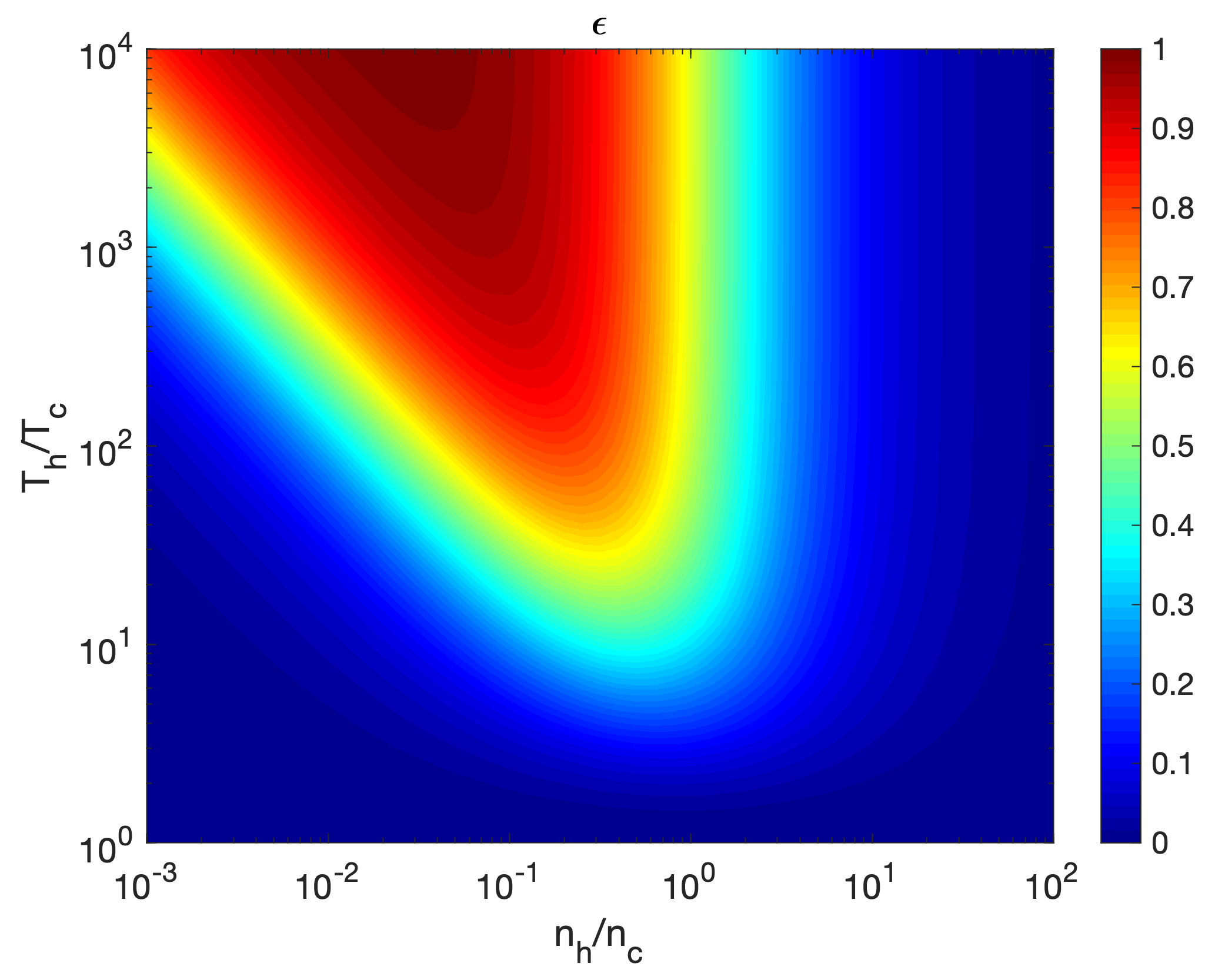}
\caption{Non-Maxwellianity $\epsilon$ as a function of $n_h/n_c$ and $T_h/T_c$ for electron distributions 
composed of two Maxwellian distributions with distinct temperatures.}
\label{epsilonTnrat}
\end{center}
\end{figure*}

To illustrate the dependence of $\epsilon$ on $n_h/n_c$ and $T_h/T_c$, we consider an electron distribution given by the sum of two stationary Maxwellian distributions with distinct densities $n_c$ and $n_h$, and distinct temperatures $T_c$ and $T_h$, where the subscripts $c$ and $h$ refer to the cold and hot distributions. 
The model Maxwellian distribition used in the calculation of $\epsilon$ has $n_e = n_c + n_h$ and temperature given by equation (\ref{Teeq}). The resulting $\epsilon$ is plotted versus $n_h/n_c$ and $T_h/T_c$ in Figure \ref{epsilonTnrat}. We find that a large region of parameter space has large values of $\epsilon$. For $n_h/n_c \lesssim 1$ and $T_h/T_c \gtrsim 10$, $\epsilon$ can reach very large values. For very large $T_h/T_c$, $\epsilon$ can approach 1. 

For Earth's outer magnetosphere $T_h/T_c$ is often $\sim 10^2$, with $T_c \sim 10$~eV and $T_c \sim 1$~keV when cold electrons are present and $n_c$ can be comparable to $n_h$ and in some cases $n_c \gg n_h$. As a result, in the outer magnetosphere $\epsilon$ will have very large values of $\epsilon$, as seen in Figure \ref{epsav2015Dec06}. 
Therefore, we expect that large values of $\epsilon$ to be observed in the magnetopause and at the magnetopause when cold electrons are present. This will result in the large values of $\epsilon$ in the magnetosphere, $n_e \lesssim 1$~cm$^{-3}$ and can account 
for the statistical results in Figure \ref{statsne}. 

\begin{figure*}[htbp!]
\begin{center}
\includegraphics[width=140mm, height=110mm]{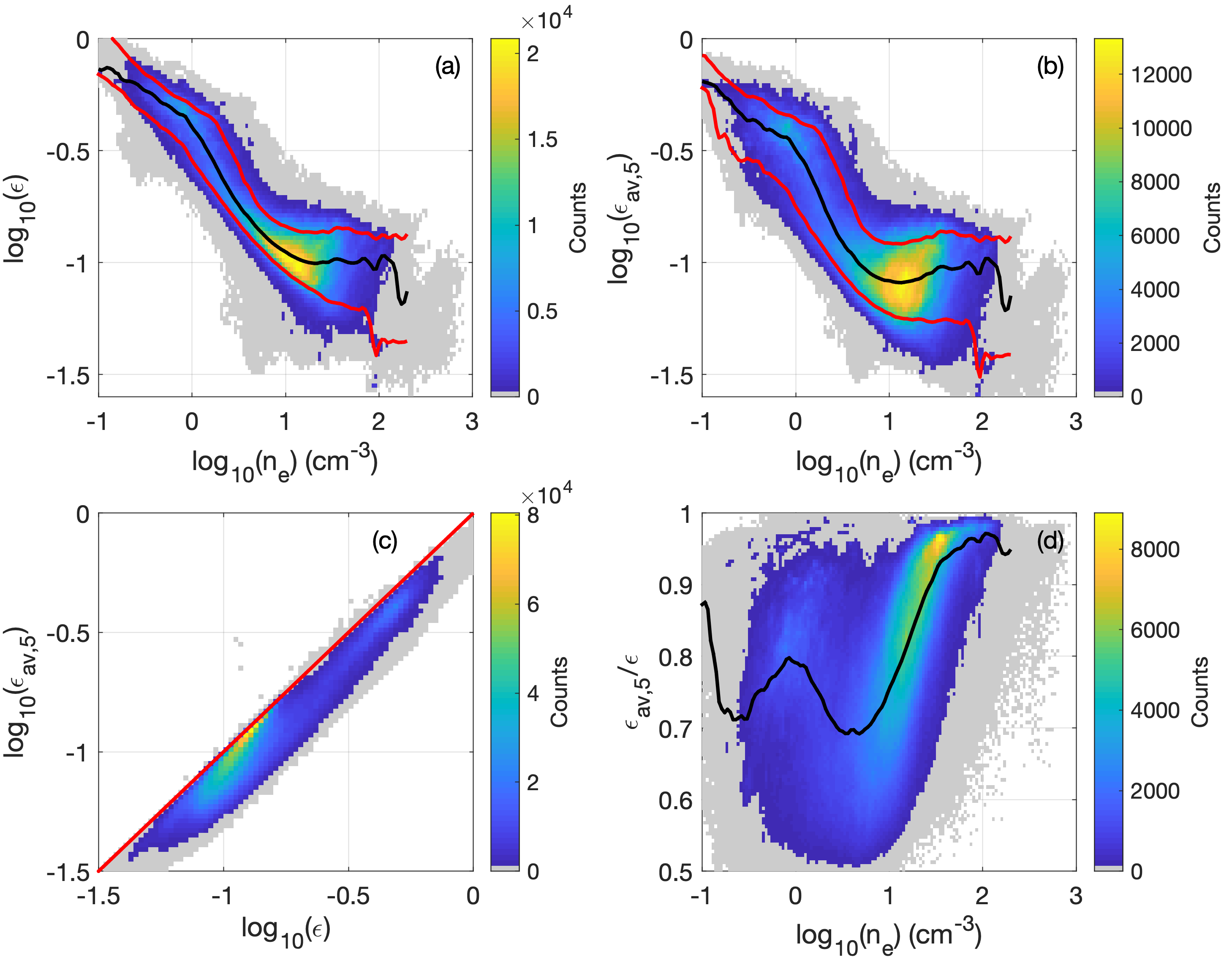}
\caption{Statstical comparison of $\epsilon$ with $\epsilon_{av,5}$. (a) Histogram of 
$\log_{10}(n_e)$ versus $\log_{10}(\epsilon)$. (b) Histogram of $\log_{10}(n_e)$ 
versus $\log_{10}(\epsilon_{av,5})$. 
The black lines in panels (a) and (b) indicate the medians (50th percentiles) of 
$\epsilon$ and $\epsilon_{av,5}$ 
as a function $n_e$ and the lower and upper red curves indicate the 10th and 90th percentiles as a function of $n_e$, respectively. 
(c) Histogram of  $\log_{10}\epsilon$ versus $\log_{10}(\epsilon_{av,5})$. The red line indicates 
$\epsilon = \epsilon_{av,5}$.
(d) Histogram of $\epsilon_{av,5}/\epsilon$ versus $n_e$. The black line is the median of  $\epsilon_{av,5}/\epsilon$ as a function of $n_e$. For direct comparison of $\epsilon$ with $\epsilon_{av,5}$ we have downsampled 
$\epsilon$ to the same sampling rate of $\epsilon_{av,5}$. }
\label{eps5avstats}
\end{center}
\end{figure*}

In Figure \ref{eps5avstats} we statistically compare $\epsilon$ with $\epsilon_{av,5}$ using the entire dataset 
used in section \ref{statisticalresults}. For direct comparison of $\epsilon$ with $\epsilon_{av,5}$ we have downsampled $\epsilon$ to the same sampling rate as $\epsilon_{av,5}$ (0.3~s or ten electron distributions). 
In Figures \ref{eps5avstats}a and \ref{eps5avstats}b we plot the histograms of $\log_{10}(\epsilon)$ and 
$\log_{10}(\epsilon_{av,5})$ versus $\log_{10}(n_e)$, respectivity. Both histograms are qualitatively very similar.  
In particular, in both plots $\epsilon$ and $\epsilon_{av,5}$ statistically decrease significantly for 
$1 \, \mathrm{cm}^{-3} \lesssim n_e \lesssim 10 \, \mathrm{cm}^{-3}$.  
The main difference is that there is a slight decrease in $\epsilon_{av,5}$ compared with $\epsilon$. 
This can be seen clearly in Figure \ref{eps5avstats}c, which plots the histogram of 
$\epsilon$ versus $\epsilon_{av,5}$. For almost all points $\epsilon_{av,5} < \epsilon$, as expected from 
averaging the observed distribution function with time. 
In Figure \ref{eps5avstats}d we plot the histogram of $\epsilon_{av,5}/\epsilon$ versus $\log_{10}(n_e)$. 
For almost all points we find that $0.5 < \epsilon_{av,5}/\epsilon < 1$, meaning that at most averaging five 64 
energy channel electron distributions reduces the non-Maxwellianity by a factor of 2. The black line 
in Figure \ref{eps5avstats}d shows the median of $\epsilon_{av,5}/\epsilon$ as a function of $n_e$. 
For $n_e \lesssim 20$~cm$^{-3}$ we find that the median remains large, $> 0.9$, indicating that averaging the electron distribution 
in time does not significantly change the results. For $n_e \gtrsim 20$~cm$^{-3}$ we find that 
the median of $\epsilon_{av,5}/\epsilon$ decreases to $0.7$. At $n_e \sim 1$~cm$^{-3}$ there is a smaller 
peak in the median at $0.8$. This $n_e$ corresponds to the typical density in the outer magnetosphere, 
where hot and cold electron distributions are common. Because the decrease in $\epsilon_{av,5}$ 
from $\epsilon$ is typically relatively small, we conclude that the increase in $\epsilon$ as $n_e$ decreases 
is physical and primarily due to the simultaneous presence of electron distributions with distinct temperatures in the outer magnetosphere.

In this Appendix we consider the effect of instrumental counting statistics on the calculated $\epsilon$. The 
key results are: 

(1) Improving the counting statistics by averaging the electron distributions over time before calculating 
$\epsilon$ results in $\epsilon$ decreasing. This decrease tends to be larger at lower densities. 
Nevertheless, $\epsilon$ increases significantly as density decreases and remains large in the outer magnetosphere and qualitatively the results do not change significantly. 

(2) The large values of $\epsilon$ in the outer magnetosphere are due to the presence of cold electrons. 
The simultaneous presence of cold and hot electron distributions with distinct temperatures is the primary 
source of non-Maxwellianity in the outer magnetosphere.

%\bibliographystyle{agufull08}
%\bibliography{magrecpapers}

\begin{thebibliography}{39}
\providecommand{\natexlab}[1]{#1}
\expandafter\ifx\csname urlstyle\endcsname\relax
  \providecommand{\doi}[1]{doi:\discretionary{}{}{}#1}\else
  \providecommand{\doi}{doi:\discretionary{}{}{}\begingroup
  \urlstyle{rm}\Url}\fi

\bibitem[{\textit{Barrie et~al.}(2019)\textit{Barrie, Cipriani, Escoubet,
  Toledo-Redondo, Nakamura, Torkar, Sternovsky, Elkington, Gershman, Giles, and
  Schiff}}]{barrie2019}
Barrie, A.~C., F.~Cipriani, C.~P. Escoubet, S.~Toledo-Redondo, R.~Nakamura,
  K.~Torkar, Z.~Sternovsky, S.~Elkington, D.~Gershman, B.~Giles, and C.~Schiff
  (2019), Characterizing spacecraft potential effects on measured particle
  trajectories, \textit{Physics of Plasmas}, \textit{26}(10), 103,504,
  \doi{10.1063/1.5119344}.

\bibitem[{\textit{{Burch} et~al.}(2016)\textit{{Burch}, {Moore}, {Torbert}, and
  {Giles}}}]{burch1}
{Burch}, J.~L., T.~E. {Moore}, R.~B. {Torbert}, and B.~L. {Giles} (2016),
  {Magnetospheric Multiscale Overview and Science Objectives}, \textit{Space
  Sci. Rev.}, \textit{199}, 5--21, \doi{10.1007/s11214-015-0164-9}.

\bibitem[{\textit{Burch et~al.}(2016)\textit{Burch, Torbert, Phan, Chen, Moore,
  Ergun, Eastwood, Gershman, Cassak, Argall, Wang, Hesse, Pollock, Giles,
  Nakamura, Mauk, Fuselier, Russell, Strangeway, Drake, Shay, Khotyaintsev,
  Lindqvist, Marklund, Wilder, Young, Torkar, Goldstein, Dorelli, Avanov, Oka,
  Baker, Jaynes, Goodrich, Cohen, Turner, Fennell, Blake, Clemmons, Goldman,
  Newman, Petrinec, Trattner, Lavraud, Reiff, Baumjohann, Magnes, Steller,
  Lewis, Saito, Coffey, and Chandler}}]{burch2}
Burch, J.~L., R.~B. Torbert, T.~D. Phan, L.-J. Chen, T.~E. Moore, R.~E. Ergun,
  J.~P. Eastwood, D.~J. Gershman, P.~A. Cassak, M.~R. Argall, S.~Wang,
  M.~Hesse, C.~J. Pollock, B.~L. Giles, R.~Nakamura, B.~H. Mauk, S.~A.
  Fuselier, C.~T. Russell, R.~J. Strangeway, J.~F. Drake, M.~A. Shay, Y.~V.
  Khotyaintsev, P.-A. Lindqvist, G.~Marklund, F.~D. Wilder, D.~T. Young,
  K.~Torkar, J.~Goldstein, J.~C. Dorelli, L.~A. Avanov, M.~Oka, D.~N. Baker,
  A.~N. Jaynes, K.~A. Goodrich, I.~J. Cohen, D.~L. Turner, J.~F. Fennell, J.~B.
  Blake, J.~Clemmons, M.~Goldman, D.~Newman, S.~M. Petrinec, K.~J. Trattner,
  B.~Lavraud, P.~H. Reiff, W.~Baumjohann, W.~Magnes, M.~Steller, W.~Lewis,
  Y.~Saito, V.~Coffey, and M.~Chandler (2016), Electron-scale measurements of
  magnetic reconnection in space, \textit{Science}, \textit{352}(6290),
  \doi{10.1126/science.aaf2939}.

\bibitem[{\textit{{Chasapis} et~al.}(2018)\textit{{Chasapis}, {Matthaeus},
  {Parashar}, {Wan}, {Haggerty}, {Pollock}, {Giles}, {Paterson}, {Dorelli},
  {Gershman}, {Torbert}, {Russell}, {Lindqvist}, {Khotyaintsev}, {Moore},
  {Ergun}, and {Burch}}}]{chasapis2018}
{Chasapis}, A., W.~H. {Matthaeus}, T.~N. {Parashar}, M.~{Wan}, C.~C.
  {Haggerty}, C.~J. {Pollock}, B.~L. {Giles}, W.~R. {Paterson}, J.~{Dorelli},
  D.~J. {Gershman}, R.~B. {Torbert}, C.~T. {Russell}, P.~A. {Lindqvist},
  Y.~{Khotyaintsev}, T.~E. {Moore}, R.~E. {Ergun}, and J.~L. {Burch} (2018),
  {In Situ Observation of Intermittent Dissipation at Kinetic Scales in the
  Earth's Magnetosheath}, \textit{Astrophysical Journal Letters},
  \textit{856}(1), L19, \doi{10.3847/2041-8213/aaadf8}.

\bibitem[{\textit{{Ergun} et~al.}(2016)\textit{{Ergun}, {Tucker}, {Westfall},
  {Goodrich}, {Malaspina}, {Summers}, {Wallace}, {Karlsson}, {Mack}, {Brennan},
  {Pyke}, {Withnell}, {Torbert}, {Macri}, {Rau}, {Dors}, {Needell},
  {Lindqvist}, {Olsson}, and {Cully}}}]{ergun3}
{Ergun}, R.~E., S.~{Tucker}, J.~{Westfall}, K.~A. {Goodrich}, D.~M.
  {Malaspina}, D.~{Summers}, J.~{Wallace}, M.~{Karlsson}, J.~{Mack},
  N.~{Brennan}, B.~{Pyke}, P.~{Withnell}, R.~{Torbert}, J.~{Macri}, D.~{Rau},
  I.~{Dors}, J.~{Needell}, P.-A. {Lindqvist}, G.~{Olsson}, and C.~M. {Cully}
  (2016), {The Axial Double Probe and Fields Signal Processing for the MMS
  Mission}, \textit{Space Sci. Rev.}, \textit{199}, 167--188,
  \doi{10.1007/s11214-014-0115-x}.

\bibitem[{\textit{Feldman et~al.}(1982)\textit{Feldman, Bame, Gary, Gosling,
  McComas, Thomsen, Paschmann, Sckopke, Hoppe, and Russell}}]{feldman1982}
Feldman, W.~C., S.~J. Bame, S.~P. Gary, J.~T. Gosling, D.~McComas, M.~F.
  Thomsen, G.~Paschmann, N.~Sckopke, M.~M. Hoppe, and C.~T. Russell (1982),
  Electron heating within the {E}arth's bow shock, \textit{Phys. Rev. Lett.},
  \textit{49}, 199--201, \doi{10.1103/PhysRevLett.49.199}.

\bibitem[{\textit{{Fuselier} et~al.}(2017)\textit{{Fuselier}, {Vines}, {Burch},
  {Petrinec}, {Trattner}, {Cassak}, {Chen}, {Ergun}, {Eriksson}, {Giles},
  {Graham}, {Khotyaintsev}, {Lavraud}, {Lewis}, {Mukherjee}, {Norgren}, {Phan},
  {Russell}, {Strangeway}, {Torbert}, and {Webster}}}]{fuselier2}
{Fuselier}, S.~A., S.~K. {Vines}, J.~L. {Burch}, S.~M. {Petrinec}, K.~J.
  {Trattner}, P.~A. {Cassak}, L.-J. {Chen}, R.~E. {Ergun}, S.~{Eriksson}, B.~L.
  {Giles}, D.~B. {Graham}, Y.~V. {Khotyaintsev}, B.~{Lavraud}, W.~S. {Lewis},
  J.~{Mukherjee}, C.~{Norgren}, T.-D. {Phan}, C.~T. {Russell}, R.~J.
  {Strangeway}, R.~B. {Torbert}, and J.~M. {Webster} (2017), {Large-scale
  characteristics of reconnection diffusion regions and associated magnetopause
  crossings observed by MMS}, \textit{Journal of Geophysical Research (Space
  Physics)}, \textit{122}, 5466--5486, \doi{10.1002/2017JA024024}.

\bibitem[{\textit{{Gary} and {Nishimura}}(2003)}]{gary5}
{Gary}, S.~P., and K.~{Nishimura} (2003), {Resonant electron firehose
  instability: Particle-in-cell simulations}, \textit{Physics of Plasmas},
  \textit{10}(9), 3571--3576, \doi{10.1063/1.1590982}.

\bibitem[{\textit{{Gary} and {Wang}}(1996)}]{gary6}
{Gary}, S.~P., and J.~{Wang} (1996), {Whistler instability: Electron anisotropy
  upper bound}, \textit{J. Geophys. Res.}, \textit{101}(A5), 10,749--10,754,
  \doi{10.1029/96JA00323}.

\bibitem[{\textit{Gary et~al.}(2012)\textit{Gary, Liu, Denton, and Wu}}]{gary2}
Gary, S.~P., K.~Liu, R.~E. Denton, and S.~Wu (2012), Whistler anisotropy
  instability with a cold electron component: {Linear theory}, \textit{J.
  Geophys. Res.}, \textit{117}, A07,203, \doi{10.1029/2012JA017631}.

\bibitem[{\textit{{Gershman} et~al.}(2017)\textit{{Gershman}, {Avanov},
  {Boardsen}, {Dorelli}, {Gliese}, {Barrie}, {Schiff}, {Paterson}, {Torbert},
  {Giles}, and {Pollock}}}]{gershman2017}
{Gershman}, D.~J., L.~A. {Avanov}, S.~A. {Boardsen}, J.~C. {Dorelli},
  U.~{Gliese}, A.~C. {Barrie}, C.~{Schiff}, W.~R. {Paterson}, R.~B. {Torbert},
  B.~L. {Giles}, and C.~J. {Pollock} (2017), {Spacecraft and Instrument
  Photoelectrons Measured by the Dual Electron Spectrometers on MMS},
  \textit{Journal of Geophysical Research (Space Physics)}, \textit{122}(A11),
  11, \doi{10.1002/2017JA024518}.

\bibitem[{\textit{Graham et~al.}(2016{\natexlab{a}})\textit{Graham,
  Khotyaintsev, Norgren, Vaivads, Andr{\'e}, Lindqvist, Marklund, Ergun,
  Paterson, Gershman, Giles, Pollock, Dorelli, Avanov, Lavraud, Saito, Magnes,
  Russell, Strangeway, Torbert, and Burch}}]{graham4}
Graham, D.~B., Y.~V. Khotyaintsev, C.~Norgren, A.~Vaivads, M.~Andr{\'e}, P.-A.
  Lindqvist, G.~T. Marklund, R.~E. Ergun, W.~R. Paterson, D.~J. Gershman, B.~L.
  Giles, C.~J. Pollock, J.~C. Dorelli, L.~A. Avanov, B.~Lavraud, Y.~Saito,
  W.~Magnes, C.~T. Russell, R.~J. Strangeway, R.~B. Torbert, and J.~L. Burch
  (2016{\natexlab{a}}), Electron currents and heating in the ion diffusion
  region of asymmetric reconnection, \textit{Geophys. Res. Lett.}, \textit{43},
  4691--4700, \doi{10.1002/2016GL068613}.

\bibitem[{\textit{Graham et~al.}(2016{\natexlab{b}})\textit{Graham, Vaivads,
  Khotyaintsev, and Andr{\'e}}}]{graham3}
Graham, D.~B., A.~Vaivads, Y.~V. Khotyaintsev, and M.~Andr{\'e}
  (2016{\natexlab{b}}), Whistler emission in the separatrix regions of
  asymmetric reconnection, \textit{J. Geophys. Res.}, \textit{121}, 1934--1954,
  \doi{10.1002/2015JA021239}.

\bibitem[{\textit{Graham et~al.}(2017)\textit{Graham, Khotyaintsev, Vaivads,
  Norgren, Andr\'e, Webster, Burch, Lindqvist, Ergun, Torbert, Paterson,
  Gershman, Giles, Magnes, and Russell}}]{graham11}
Graham, D.~B., Y.~V. Khotyaintsev, A.~Vaivads, C.~Norgren, M.~Andr\'e, J.~M.
  Webster, J.~L. Burch, P.-A. Lindqvist, R.~E. Ergun, R.~B. Torbert, W.~R.
  Paterson, D.~J. Gershman, B.~L. Giles, W.~Magnes, and C.~T. Russell (2017),
  Instability of agyrotropic electron beams near the electron diffusion region,
  \textit{Phys. Rev. Lett.}, \textit{119}, 025,101,
  \doi{10.1103/PhysRevLett.119.025101}.

\bibitem[{\textit{Graham et~al.}(2018)\textit{Graham, Vaivads, Khotyaintsev,
  Eriksson, Andre, Malaspina, Lindqvist, Gershman, and Plaschke}}]{graham2018b}
Graham, D.~B., A.~Vaivads, Y.~V. Khotyaintsev, A.~I. Eriksson, M.~Andre, D.~M.
  Malaspina, P.-A. Lindqvist, D.~J. Gershman, and F.~Plaschke (2018), Enhanced
  escape of spacecraft photoelectrons caused by {L}angmuir and upper hybrid
  waves, \textit{Journal of Geophysical Research: Space Physics},
  \textit{123}(9), 7534--7553, \doi{10.1029/2018JA025874}.

\bibitem[{\textit{{Greco} et~al.}(2012)\textit{{Greco}, {Valentini},
  {Servidio}, and {Matthaeus}}}]{greco2012}
{Greco}, A., F.~{Valentini}, S.~{Servidio}, and W.~H. {Matthaeus} (2012),
  {Inhomogeneous kinetic effects related to intermittent magnetic
  discontinuities}, \textit{Phys. Rev. E}, \textit{86}(6), 066,405,
  \doi{10.1103/PhysRevE.86.066405}.

\bibitem[{\textit{Kennel and Petschek}(1966)}]{kennel1}
Kennel, C.~F., and H.~E. Petschek (1966), Limit on stably trapped particle
  fluxes, \textit{J. Geophys. Res.}, \textit{71}, 1--28,
  \doi{10.1029/JZ071i001p00001}.

\bibitem[{\textit{Khotyaintsev et~al.}(2016)\textit{Khotyaintsev, Graham,
  Norgren, Eriksson, Li, Johlander, Vaivads, Andr{\'e}, Pritchett, Retino,
  Phan, Ergun, Goodrich, Lindqvist, Marklund, Contel, Plaschke, Magnes,
  Strangeway, Russell, Vaith, Argall, Kletzing, Nakamura, Torbert, Paterson,
  Gershman, Dorelli, Avanov, Lavraud, Saito, Giles, Pollock, Turner, Blake,
  Fennell, Jaynes, Mauk, and Burch}}]{khotyaintsev4}
Khotyaintsev, Y.~V., D.~B. Graham, C.~Norgren, E.~Eriksson, W.~Li,
  A.~Johlander, A.~Vaivads, M.~Andr{\'e}, P.~L. Pritchett, A.~Retino, T.~D.
  Phan, R.~E. Ergun, K.~Goodrich, P.-A. Lindqvist, G.~T. Marklund, O.~L.
  Contel, F.~Plaschke, W.~Magnes, R.~J. Strangeway, C.~T. Russell, H.~Vaith,
  M.~R. Argall, C.~A. Kletzing, R.~Nakamura, R.~B. Torbert, W.~R. Paterson,
  D.~J. Gershman, J.~C. Dorelli, L.~A. Avanov, B.~Lavraud, Y.~Saito, B.~L.
  Giles, C.~J. Pollock, D.~L. Turner, J.~D. Blake, J.~F. Fennell, A.~Jaynes,
  B.~H. Mauk, and J.~L. Burch (2016), Electron jet of asymmetric reconnection,
  \textit{Geophys. Res. Lett.}, \textit{43}, 5571--5580,
  \doi{10.1002/2016GL069064}.

\bibitem[{\textit{Khotyaintsev et~al.}(2019)\textit{Khotyaintsev, Graham,
  Norgren, and Vaivads}}]{khotyaintsev2019}
Khotyaintsev, Y.~V., D.~B. Graham, C.~Norgren, and A.~Vaivads (2019),
  Collisionless magnetic reconnection and waves: Progress review,
  \textit{Frontiers in Astronomy and Space Sciences}, \textit{6}, 70,
  \doi{10.3389/fspas.2019.00070}.

\bibitem[{\textit{Li and Habbal}(2000)}]{li3}
Li, X., and S.~R. Habbal (2000), Electron kinetic firehose instability,
  \textit{Journal of Geophysical Research: Space Physics}, \textit{105}(A12),
  27,377--27,385, \doi{10.1029/2000JA000063}.

\bibitem[{\textit{Liang et~al.}(2020)\textit{Liang, Barbhuiya, Cassak, Pezzi,
  Servidio, Valentini, and Zank}}]{liang2020}
Liang, H., M.~H. Barbhuiya, P.~A. Cassak, O.~Pezzi, S.~Servidio, F.~Valentini,
  and G.~P. Zank (2020), Kinetic entropy-based measures of distribution
  function non-maxwellianity: theory and simulations, \textit{Journal of Plasma
  Physics}, \textit{86}(5), 825860,502, \doi{10.1017/S0022377820001270}.

\bibitem[{\textit{{Lindqvist} et~al.}(2016)\textit{{Lindqvist}, {Olsson},
  {Torbert}, {King}, {Granoff}, {Rau}, {Needell}, {Turco}, {Dors}, {Beckman},
  {Macri}, {Frost}, {Salwen}, {Eriksson}, {{\AA}hl{\'e}n}, {Khotyaintsev},
  {Porter}, {Lappalainen}, {Ergun}, {Wermeer}, and {Tucker}}}]{lindqvist1}
{Lindqvist}, P.-A., G.~{Olsson}, R.~B. {Torbert}, B.~{King}, M.~{Granoff},
  D.~{Rau}, G.~{Needell}, S.~{Turco}, I.~{Dors}, P.~{Beckman}, J.~{Macri},
  C.~{Frost}, J.~{Salwen}, A.~{Eriksson}, L.~{{\AA}hl{\'e}n}, Y.~V.
  {Khotyaintsev}, J.~{Porter}, K.~{Lappalainen}, R.~E. {Ergun}, W.~{Wermeer},
  and S.~{Tucker} (2016), {The Spin-Plane Double Probe Electric Field
  Instrument for MMS}, \textit{Space Sci. Rev.}, \textit{199}, 137--165,
  \doi{10.1007/s11214-014-0116-9}.

\bibitem[{\textit{{Norgren} et~al.}(2016)\textit{{Norgren}, {Graham},
  {Khotyaintsev}, {Andr{\'e}}, {Vaivads}, {Chen}, {Lindqvist}, {Marklund},
  {Ergun}, {Magnes}, {Strangeway}, {Russell}, {Torbert}, {Paterson},
  {Gershman}, {Dorelli}, {Avanov}, {Lavraud}, {Saito}, {Giles}, {Pollock}, and
  {Burch}}}]{norgren4}
{Norgren}, C., D.~B. {Graham}, Y.~V. {Khotyaintsev}, M.~{Andr{\'e}},
  A.~{Vaivads}, L.-J. {Chen}, P.-A. {Lindqvist}, G.~T. {Marklund}, R.~E.
  {Ergun}, W.~{Magnes}, R.~J. {Strangeway}, C.~T. {Russell}, R.~B. {Torbert},
  W.~R. {Paterson}, D.~J. {Gershman}, J.~C. {Dorelli}, L.~A. {Avanov},
  B.~{Lavraud}, Y.~{Saito}, B.~L. {Giles}, C.~J. {Pollock}, and J.~L. {Burch}
  (2016), Finite gyroradius effects in the electron outflow of asymmetric
  magnetic reconnection, \textit{Geophys. Res. Lett.}, \textit{43}, 6724--6733,
  \doi{10.1002/2016GL069205}.

\bibitem[{\textit{Oka et~al.}(2017)\textit{Oka, {Wilson III}, Phan, Hull,
  Amano, Hoshino, Argall, Contel, Agapitov, Gershman, Khotyaintsev, Burch,
  Torbert, Pollock, Dorelli, Giles, Moore, Saito, Avanov, Paterson, Ergun,
  Strangeway, Russell, and Lindqvist}}]{oka2017}
Oka, M., L.~B. {Wilson III}, T.~D. Phan, A.~J. Hull, T.~Amano, M.~Hoshino,
  M.~R. Argall, O.~L. Contel, O.~Agapitov, D.~J. Gershman, Y.~V. Khotyaintsev,
  J.~L. Burch, R.~B. Torbert, C.~Pollock, J.~C. Dorelli, B.~L. Giles, T.~E.
  Moore, Y.~Saito, L.~A. Avanov, W.~Paterson, R.~E. Ergun, R.~J. Strangeway,
  C.~T. Russell, and P.~A. Lindqvist (2017), Electron scattering by
  high-frequency whistler waves at {E}arth's bow shock, \textit{The
  Astrophysical Journal}, \textit{842}(2), L11, \doi{10.3847/2041-8213/aa7759}.

\bibitem[{\textit{Perri et~al.}(2020)\textit{Perri, Perrone, Yordanova,
  Sorriso-Valvo, Paterson, Gershman, Giles, Pollock, Dorelli, Avanov, and
  et~al.}}]{perri2020}
Perri, S., D.~Perrone, E.~Yordanova, L.~Sorriso-Valvo, W.~R. Paterson, D.~J.
  Gershman, B.~L. Giles, C.~J. Pollock, J.~C. Dorelli, L.~A. Avanov, and et~al.
  (2020), On the deviation from maxwellian of the ion velocity distribution
  functions in the turbulent magnetosheath, \textit{Journal of Plasma Physics},
  \textit{86}(1), 905860,108, \doi{10.1017/S0022377820000021}.

\bibitem[{\textit{Phan et~al.}(2016)\textit{Phan, Eastwood, Cassak, Øieroset,
  Gosling, Gershman, Mozer, Shay, Fujimoto, Daughton, Drake, Burch, Torbert,
  Ergun, Chen, Wang, Pollock, Dorelli, Lavraud, Giles, Moore, Saito, Avanov,
  Paterson, Strangeway, Russell, Khotyaintsev, Lindqvist, Oka, and
  Wilder}}]{phan3}
Phan, T.~D., J.~P. Eastwood, P.~A. Cassak, M.~Øieroset, J.~T. Gosling, D.~J.
  Gershman, F.~S. Mozer, M.~A. Shay, M.~Fujimoto, W.~Daughton, J.~F. Drake,
  J.~L. Burch, R.~B. Torbert, R.~E. Ergun, L.~J. Chen, S.~Wang, C.~Pollock,
  J.~C. Dorelli, B.~Lavraud, B.~L. Giles, T.~E. Moore, Y.~Saito, L.~A. Avanov,
  W.~Paterson, R.~J. Strangeway, C.~T. Russell, Y.~Khotyaintsev, P.~A.
  Lindqvist, M.~Oka, and F.~D. Wilder (2016), {MMS} observations of
  electron-scale filamentary currents in the reconnection exhaust and near the
  {X} line, \textit{Geophysical Research Letters}, \textit{43}(12), 6060--6069,
  \doi{10.1002/2016GL069212}.

\bibitem[{\textit{{Pollock} et~al.}(2016)\textit{{Pollock}, {Moore}, {Jacques},
  {Burch}, {Gliese}, {Saito}, {Omoto}, {Avanov}, {Barrie}, {Coffey}, {Dorelli},
  {Gershman}, {Giles}, {Rosnack}, {Salo}, {Yokota}, {Adrian}, {Aoustin},
  {Auletti}, {Aung}, {Bigio}, {Cao}, {Chandler}, {Chornay}, {Christian},
  {Clark}, {Collinson}, {Corris}, {De Los Santos}, {Devlin}, {Diaz},
  {Dickerson}, {Dickson}, {Diekmann}, {Diggs}, {Duncan}, {Figueroa-Vinas},
  {Firman}, {Freeman}, {Galassi}, {Garcia}, {Goodhart}, {Guererro}, {Hageman},
  {Hanley}, {Hemminger}, {Holland}, {Hutchins}, {James}, {Jones}, {Kreisler},
  {Kujawski}, {Lavu}, {Lobell}, {LeCompte}, {Lukemire}, {MacDonald}, {Mariano},
  {Mukai}, {Narayanan}, {Nguyan}, {Onizuka}, {Paterson}, {Persyn}, {Piepgrass},
  {Cheney}, {Rager}, {Raghuram}, {Ramil}, {Reichenthal}, {Rodriguez},
  {Rouzaud}, {Rucker}, {Saito}, {Samara}, {Sauvaud}, {Schuster}, {Shappirio},
  {Shelton}, {Sher}, {Smith}, {Smith}, {Smith}, {Steinfeld}, {Szymkiewicz},
  {Tanimoto}, {Taylor}, {Tucker}, {Tull}, {Uhl}, {Vloet}, {Walpole}, {Weidner},
  {White}, {Winkert}, {Yeh}, and {Zeuch}}}]{pollock1}
{Pollock}, C., T.~{Moore}, A.~{Jacques}, J.~{Burch}, U.~{Gliese}, Y.~{Saito},
  T.~{Omoto}, L.~{Avanov}, A.~{Barrie}, V.~{Coffey}, J.~{Dorelli},
  D.~{Gershman}, B.~{Giles}, T.~{Rosnack}, C.~{Salo}, S.~{Yokota}, M.~{Adrian},
  C.~{Aoustin}, C.~{Auletti}, S.~{Aung}, V.~{Bigio}, N.~{Cao}, M.~{Chandler},
  D.~{Chornay}, K.~{Christian}, G.~{Clark}, G.~{Collinson}, T.~{Corris}, A.~{De
  Los Santos}, R.~{Devlin}, T.~{Diaz}, T.~{Dickerson}, C.~{Dickson},
  A.~{Diekmann}, F.~{Diggs}, C.~{Duncan}, A.~{Figueroa-Vinas}, C.~{Firman},
  M.~{Freeman}, N.~{Galassi}, K.~{Garcia}, G.~{Goodhart}, D.~{Guererro},
  J.~{Hageman}, J.~{Hanley}, E.~{Hemminger}, M.~{Holland}, M.~{Hutchins},
  T.~{James}, W.~{Jones}, S.~{Kreisler}, J.~{Kujawski}, V.~{Lavu}, J.~{Lobell},
  E.~{LeCompte}, A.~{Lukemire}, E.~{MacDonald}, A.~{Mariano}, T.~{Mukai},
  K.~{Narayanan}, Q.~{Nguyan}, M.~{Onizuka}, W.~{Paterson}, S.~{Persyn},
  B.~{Piepgrass}, F.~{Cheney}, A.~{Rager}, T.~{Raghuram}, A.~{Ramil},
  L.~{Reichenthal}, H.~{Rodriguez}, J.~{Rouzaud}, A.~{Rucker}, Y.~{Saito},
  M.~{Samara}, J.-A. {Sauvaud}, D.~{Schuster}, M.~{Shappirio}, K.~{Shelton},
  D.~{Sher}, D.~{Smith}, K.~{Smith}, S.~{Smith}, D.~{Steinfeld},
  R.~{Szymkiewicz}, K.~{Tanimoto}, J.~{Taylor}, C.~{Tucker}, K.~{Tull},
  A.~{Uhl}, J.~{Vloet}, P.~{Walpole}, S.~{Weidner}, D.~{White}, G.~{Winkert},
  P.-S. {Yeh}, and M.~{Zeuch} (2016), {Fast Plasma Investigation for
  Magnetospheric Multiscale}, \textit{Space Sci. Rev.}, \textit{199}, 331--406,
  \doi{10.1007/s11214-016-0245-4}.

\bibitem[{\textit{{Russell} et~al.}(2016)\textit{{Russell}, {Anderson},
  {Baumjohann}, {Bromund}, {Dearborn}, {Fischer}, {Le}, {Leinweber}, {Leneman},
  {Magnes}, {Means}, {Moldwin}, {Nakamura}, {Pierce}, {Plaschke}, {Rowe},
  {Slavin}, {Strangeway}, {Torbert}, {Hagen}, {Jernej}, {Valavanoglou}, and
  {Richter}}}]{russell2}
{Russell}, C.~T., B.~J. {Anderson}, W.~{Baumjohann}, K.~R. {Bromund},
  D.~{Dearborn}, D.~{Fischer}, G.~{Le}, H.~K. {Leinweber}, D.~{Leneman},
  W.~{Magnes}, J.~D. {Means}, M.~B. {Moldwin}, R.~{Nakamura}, D.~{Pierce},
  F.~{Plaschke}, K.~M. {Rowe}, J.~A. {Slavin}, R.~J. {Strangeway},
  R.~{Torbert}, C.~{Hagen}, I.~{Jernej}, A.~{Valavanoglou}, and I.~{Richter}
  (2016), The magnetospheric multiscale magnetometers, \textit{Space Sci.
  Rev.}, \textit{199}, 189--256, \doi{10.1007/s11214-014-0057-3}.

\bibitem[{\textit{{Scudder}}(1995)}]{scudder1995}
{Scudder}, J.~D. (1995), {A review of the physics of electron heating at
  collisionless shocks}, \textit{Advances in Space Research}, \textit{15},
  181--223, \doi{10.1016/0273-1177(94)00101-6}.

\bibitem[{\textit{Servidio et~al.}(2017)\textit{Servidio, Chasapis, Matthaeus,
  Perrone, Valentini, Parashar, Veltri, Gershman, Russell, Giles, Fuselier,
  Phan, and Burch}}]{servidio2017}
Servidio, S., A.~Chasapis, W.~H. Matthaeus, D.~Perrone, F.~Valentini, T.~N.
  Parashar, P.~Veltri, D.~Gershman, C.~T. Russell, B.~Giles, S.~A. Fuselier,
  T.~D. Phan, and J.~Burch (2017), Magnetospheric multiscale observation of
  plasma velocity-space cascade: Hermite representation and theory,
  \textit{Phys. Rev. Lett.}, \textit{119}, 205,101,
  \doi{10.1103/PhysRevLett.119.205101}.

\bibitem[{\textit{{Swisdak}}(2016)}]{swisdak2}
{Swisdak}, M. (2016), Quantifying gyrotropy in magnetic reconnection,
  \textit{Geophys. Res. Lett.}, \textit{43}, 43--49,
  \doi{10.1002/2015GL066980}.

\bibitem[{\textit{Toledo-Redondo et~al.}(2016)\textit{Toledo-Redondo,
  Andr{\'e}, Khotyaintsev, Vaivads, Walsh, Li, Graham, Lavraud, Masson, Aunai,
  Divin, Dargent, Fuselier, Gershman, Dorelli, Giles, Avanov, Pollock, Saito,
  Moore, Coffey, Chandler, Lindqvist, Torbert, and Russell}}]{toledoredondo2}
Toledo-Redondo, S., M.~Andr{\'e}, Y.~V. Khotyaintsev, A.~Vaivads, A.~Walsh,
  W.~Li, D.~B. Graham, B.~Lavraud, A.~Masson, N.~Aunai, A.~Divin, J.~Dargent,
  S.~Fuselier, D.~J. Gershman, J.~Dorelli, B.~Giles, L.~Avanov, C.~Pollock,
  Y.~Saito, T.~E. Moore, V.~Coffey, M.~O. Chandler, P.-A. Lindqvist,
  R.~Torbert, and C.~T. Russell (2016), Cold ion demagnetization near the
  {X}-line of magnetic reconnection, \textit{Geophysical Research Letters},
  \textit{43}(13), 6759--6767.

\bibitem[{\textit{Toledo-Redondo et~al.}(2019)\textit{Toledo-Redondo, Lavraud,
  Fuselier, André, Khotyaintsev, Nakamura, Escoubet, Li, Torkar, Cipriani,
  Barrie, Giles, Moore, Gershman, Lindqvist, Ergun, Russell, and
  Burch}}]{toledoredondo2019}
Toledo-Redondo, S., B.~Lavraud, S.~A. Fuselier, M.~André, Y.~V. Khotyaintsev,
  R.~Nakamura, C.~P. Escoubet, W.~Y. Li, K.~Torkar, F.~Cipriani, A.~C. Barrie,
  B.~Giles, T.~E. Moore, D.~Gershman, P.-A. Lindqvist, R.~E. Ergun, C.~T.
  Russell, and J.~L. Burch (2019), Electrostatic spacecraft potential structure
  and wake formation effects for characterization of cold ion beams in the
  {E}arth's magnetosphere, \textit{Journal of Geophysical Research: Space
  Physics}, \textit{124}(12), 10,048--10,062, \doi{10.1029/2019JA027145}.

\bibitem[{\textit{Torkar et~al.}(2016)\textit{Torkar, Nakamura, Tajmar,
  Scharlemann, Jeszenszky, Laky, Fremuth, Escoubet, and Svenes}}]{Torkar2016}
Torkar, K., R.~Nakamura, M.~Tajmar, C.~Scharlemann, H.~Jeszenszky, G.~Laky,
  G.~Fremuth, C.~P. Escoubet, and K.~Svenes (2016), Active spacecraft potential
  control investigation, \textit{Space Science Reviews}, \textit{199},
  515--544, \doi{10.1007/s11214-014-0049-3}.

\bibitem[{\textit{{Valentini} et~al.}(2016)\textit{{Valentini}, {Perrone},
  {Stabile}, {Pezzi}, {Servidio}, {De Marco}, {Marcucci}, {Bruno}, {Lavraud},
  {De Keyser}, {Consolini}, {Brienza}, {Sorriso-Valvo}, {Retin{\`o}},
  {Vaivads}, {Salatti}, and {Veltri}}}]{valentini2016}
{Valentini}, F., D.~{Perrone}, S.~{Stabile}, O.~{Pezzi}, S.~{Servidio}, R.~{De
  Marco}, F.~{Marcucci}, R.~{Bruno}, B.~{Lavraud}, J.~{De Keyser},
  G.~{Consolini}, D.~{Brienza}, L.~{Sorriso-Valvo}, A.~{Retin{\`o}},
  A.~{Vaivads}, M.~{Salatti}, and P.~{Veltri} (2016), {Differential kinetic
  dynamics and heating of ions in the turbulent solar wind}, \textit{New
  Journal of Physics}, \textit{18}, 125,001,
  \doi{10.1088/1367-2630/18/12/125001}.

\bibitem[{\textit{Voros et~al.}(2017)\textit{Voros, Yordanova, Varsani,
  Genestreti, Khotyaintsev, Li, Graham, Norgren, Nakamura, Narita, Plaschke,
  Magnes, Baumjohann, Fischer, Vaivads, Eriksson, Lindqvist, Marklund, Ergun,
  Leitner, Leubner, Strangeway, Le~Contel, Pollock, Giles, Torbert, Burch,
  Avanov, Dorelli, Gershman, Paterson, Lavraud, and Saito}}]{voros2017}
Voros, Z., E.~Yordanova, A.~Varsani, K.~J. Genestreti, Y.~V. Khotyaintsev,
  W.~Li, D.~B. Graham, C.~Norgren, R.~Nakamura, Y.~Narita, F.~Plaschke,
  W.~Magnes, W.~Baumjohann, D.~Fischer, A.~Vaivads, E.~Eriksson, P.-A.
  Lindqvist, G.~Marklund, R.~E. Ergun, M.~Leitner, M.~P. Leubner, R.~J.
  Strangeway, O.~Le~Contel, C.~Pollock, B.~J. Giles, R.~B. Torbert, J.~L.
  Burch, L.~A. Avanov, J.~C. Dorelli, D.~J. Gershman, W.~R. Paterson,
  B.~Lavraud, and Y.~Saito (2017), {MMS} observation of magnetic reconnection
  in the turbulent magnetosheath, \textit{Journal of Geophysical Research:
  Space Physics}, \textit{122}, 11,442--11,467, \doi{10.1002/2017JA024535}.

\bibitem[{\textit{Walsh et~al.}(2020)\textit{Walsh, Hull, Agapitov, Mozer, and
  Li}}]{walsh2020}
Walsh, B.~M., A.~J. Hull, O.~Agapitov, F.~S. Mozer, and H.~Li (2020), A census
  of magnetospheric electrons from several e{V} to 30 ke{V}, \textit{Journal of
  Geophysical Research: Space Physics}, \textit{125}(5), e2019JA027,577,
  \doi{10.1029/2019JA027577}.

\bibitem[{\textit{{Webster} et~al.}(2018)\textit{{Webster}, {Burch}, {Reiff},
  {Daou}, {Genestreti}, {Graham}, {Torbert}, {Ergun}, {Sazykin}, {Marshall},
  {Allen}, {Chen}, {Wang}, {Phan}, {Giles}, {Moore}, {Fuselier}, {Cozzani},
  {Russell}, {Eriksson}, {Rager}, {Broll}, {Goodrich}, and
  {Wilder}}}]{webster1}
{Webster}, J.~M., J.~L. {Burch}, P.~H. {Reiff}, A.~G. {Daou}, K.~J.
  {Genestreti}, D.~B. {Graham}, R.~B. {Torbert}, R.~E. {Ergun}, S.~Y.
  {Sazykin}, A.~{Marshall}, R.~C. {Allen}, L.-J. {Chen}, S.~{Wang}, T.~D.
  {Phan}, B.~L. {Giles}, T.~E. {Moore}, S.~A. {Fuselier}, G.~{Cozzani}, C.~T.
  {Russell}, S.~{Eriksson}, A.~C. {Rager}, J.~M. {Broll}, K.~{Goodrich}, and
  F.~{Wilder} (2018), {Magnetospheric Multiscale Dayside Reconnection Electron
  Diffusion Region Events}, \textit{Journal of Geophysical Research (Space
  Physics)}, \textit{123}, 4858--4878, \doi{10.1029/2018JA025245}.

\bibitem[{\textit{Yordanova et~al.}(2016)\textit{Yordanova, Voros, Varsani,
  Graham, Norgren, Khotyaintsev, Vaivads, Eriksson, Nakamura, Lindqvist,
  Marklund, Ergun, Magnes, Baumjohann, Fischer, Plaschke, Narita, Russell,
  Strangeway, Le~Contel, Pollock, Torbert, Giles, Burch, Avanov, Dorelli,
  Gershman, Paterson, Lavraud, and Saito}}]{yordanova2016}
Yordanova, E., Z.~Voros, A.~Varsani, D.~B. Graham, C.~Norgren, Y.~V.
  Khotyaintsev, A.~Vaivads, E.~Eriksson, R.~Nakamura, P.-A. Lindqvist,
  G.~Marklund, R.~E. Ergun, W.~Magnes, W.~Baumjohann, D.~Fischer, F.~Plaschke,
  Y.~Narita, C.~T. Russell, R.~J. Strangeway, O.~Le~Contel, C.~Pollock, R.~B.
  Torbert, B.~J. Giles, J.~L. Burch, L.~A. Avanov, J.~C. Dorelli, D.~J.
  Gershman, W.~R. Paterson, B.~Lavraud, and Y.~Saito (2016), Electron scale
  structures and magnetic reconnection signatures in the turbulent
  magnetosheath, \textit{Geophys. Res. Lett.}, \textit{43}(12), 5969--5978,
  \doi{10.1002/2016GL069191}.

\end{thebibliography}

\end{document}